\documentclass[]{aa}

\usepackage{txfonts}
\usepackage{graphicx}
\usepackage{natbib}
\bibpunct{(}{)}{;}{a}{}{,}

\newcommand{\modified}[1]{{{#1}}}

\title{Faraday Rotation Measure Synthesis}
\titlerunning{RM-synthesis}

\author{M.~A. Brentjens\inst{1}\fnmsep\inst{2}\and 
        A.~G. de Bruyn\inst{2}\fnmsep\inst{1}}

\authorrunning{Brentjens \and de Bruyn}
\institute{Kapteyn Astronomical Institute, University of Groningen,
P.O. Box 800, 9700 AV Groningen, the Netherlands \and ASTRON, P.O. Box
2, 7990 AA Dwingeloo, the Netherlands}

\date{Received 07 March 2005 / Accepted 20 June 2005}

\abstract{We extend the rotation measure work of \citet{Burn1966} to
the cases of limited sampling of $\lambda^2$ space and non-constant
emission spectra. We introduce the rotation measure transfer function
(RMTF), which is an excellent predictor of $n\pi$ ambiguity problems
with the $\lambda^2$ coverage. Rotation measure synthesis can be
implemented very efficiently on modern computers. Because the analysis
is easily applied to wide fields, one can conduct very fast RM surveys
of weak spatially extended sources. Difficult situations, for example
multiple sources along the line of sight, are easily detected and
transparently handled. Under certain conditions, it is even possible
to recover the emission as a function of Faraday depth within a single
cloud of ionized gas. Rotation measure synthesis has already been
successful in discovering widespread, weak, polarized emission
associated with the Perseus cluster \citep{DeBruynBrentjens2005}. In
simple, high signal to noise situations it is as good as traditional
linear fits to $\chi$ versus $\lambda^2$ plots. However, when the
situation is more complex or very weak polarized emission at high
rotation measures is expected, it is the only viable option.
}

\begin{document}
\maketitle

\section{Introduction}
\label{brentjens_sec:introduction}

Polarization observations at radio frequencies are an important
diagnostic tool in the study of galactic and extragalactic magnetic
fields \citep[e.g.][]{Kronberg1994,Vallee1997,Widrow2002}. Due to
birefringence of the magneto-ionic medium, the polarization angle of
linearly polarized radiation that propagates through the plasma is
rotated as a function of frequency. This effect is called Faraday
rotation. There exist many papers describing aspects of Faraday
rotation work. The most relevant ones for this work are
\citet{Burn1966}, \citet{GardnerWhiteoak1966}, \citet{SokoloffEtAl1998},
\citet{SokoloffEtAl1999}, and \citet{Vallee1980}.

Assuming that the directions of the velocity vectors of the
electrons gyrating in a magnetized plasma are isotropically distributed,
\citet{LeRoux1961} showed that the intrinsic degree of polarization of
synchrotron radiation from plasma in a uniform magnetic field
is given by 
\begin{equation}
\|p\| = \frac{3\gamma + 3}{3\gamma+7},
\label{brentjens_eqn:polpercentage}
\end{equation}
independent of frequency and viewing angle. In this equation, $\gamma$
is the spectral index of the relativistic electron
distribution in energy
\begin{equation}
n_\mathrm{e}(E)\ \mathrm{d}E = AE^{-\gamma}\ \mathrm{d}E,
\end{equation}
where $n_e(E)\mathrm{d}E$ is the density of the electrons between
energies $E$ and $E+\mathrm{d}E$. The density of electrons
having energies between 1 and $1+\mathrm{d}E$ is $A\mathrm{d}E$. The
total electron density $n_\mathrm{e} =
\int_\mathrm{E_0}^\infty n_\mathrm{e}(E) \mathrm{d}E$, where $E_0$ is
a cutoff energy that is required in order to let the integral
converge. 

From observations of the Crab nebula by \citet{Woltjer1958},
\citet{Westfold1959} determined that $\gamma \approx
\frac{5}{3}$. This would imply a polarization fraction of 
approximately 67\%, independent of frequency. In many radio sources,
the observed polarization fractions are much lower. Usually the
polarization fraction decreases steeply with increasing wavelength
\citep{ConwayStrom1985,StromConway1985}.

\citet{Burn1966} discusses this depolarization effect
extensively. One of the mechanisms he discusses is Faraday dispersion:
emission at different Faraday depths along the same line of sight.

\begin{figure*}
\centering
\begin{minipage}{0.495\textwidth}
\includegraphics[width=\textwidth]{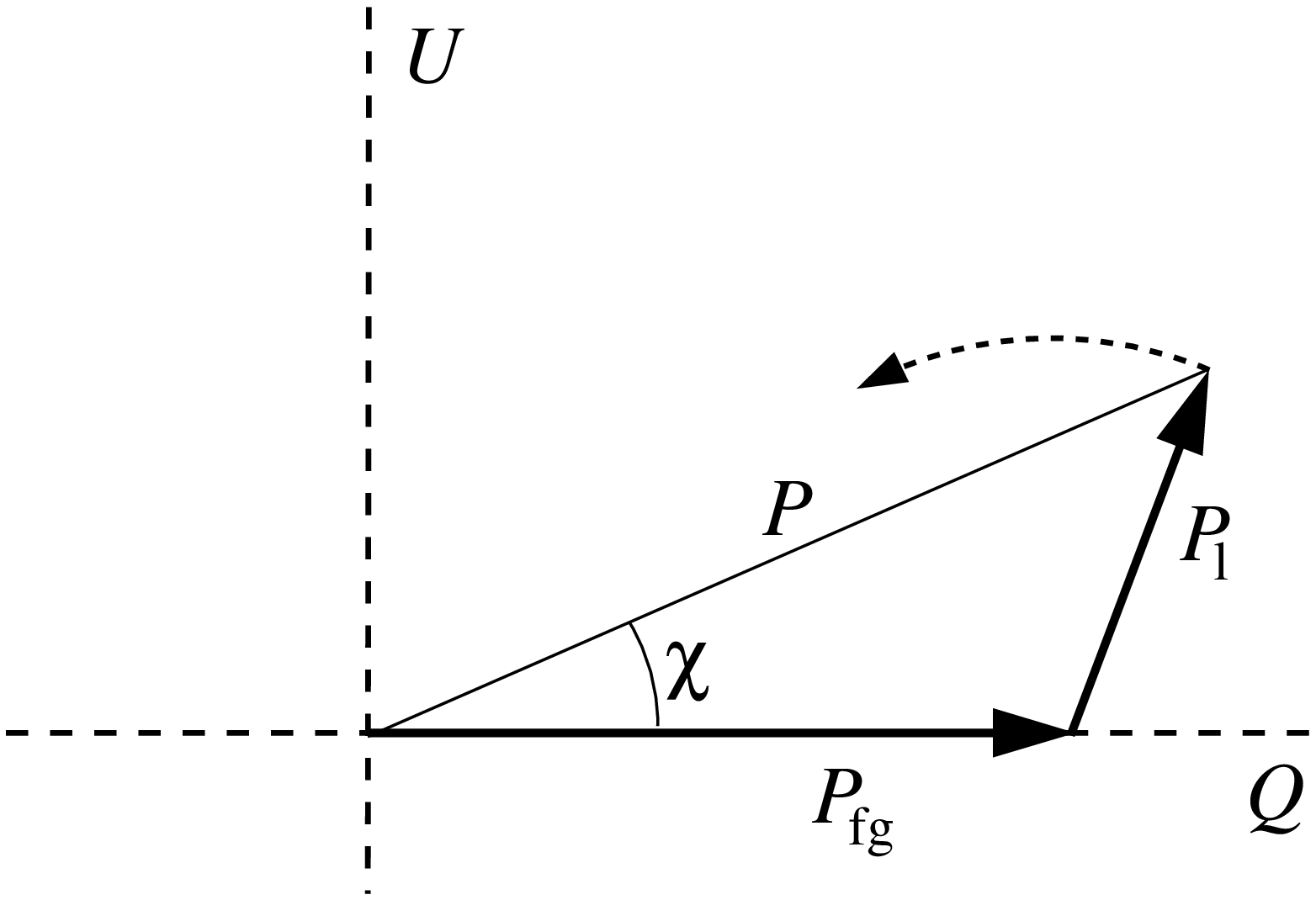}
\end{minipage}
\begin{minipage}{0.495\textwidth}
\includegraphics[width=\textwidth]{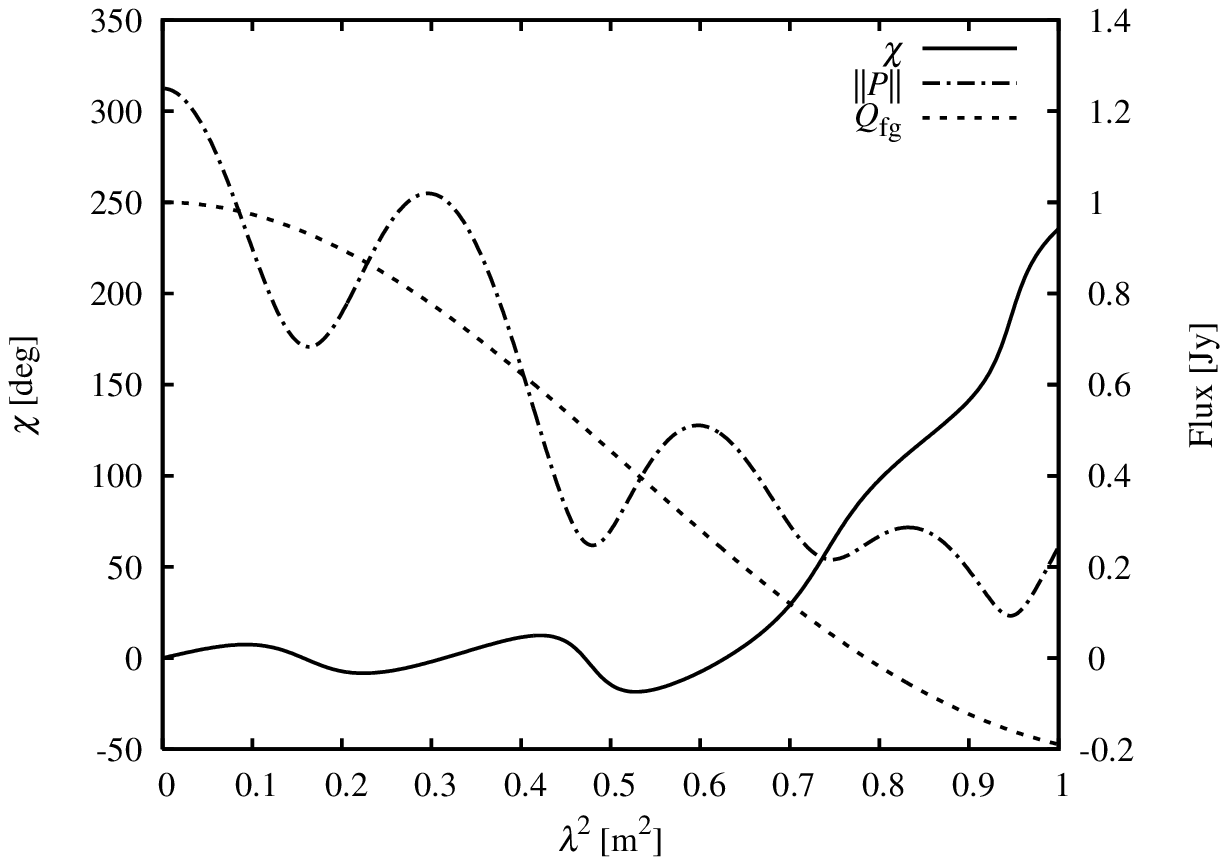}
\end{minipage}
\caption{$P(\lambda^2)$ for a simple, yet non-trivial Faraday
dispersion function: a synchrotron emitting and Faraday rotating slab
at $-2 \le \phi \le +2\ \mbox{rad}\ \mbox{m}^{-2}$ (the Galactic
foreground) and a Faraday thin source at $\phi = +10\ \mbox{rad}\
\mbox{m}^{-2}$ (a distant radio lobe). The lefthand panel shows
qualitatively how the individual ($Q$, $U$) vectors of the foreground and
the lobe relate to the total polarization $\|P\|$ and the polarization
angle $\chi$. It reflects the situation around $\lambda^2\approx
0.38$~m$^{2}$. $P_\mathrm{l}$ rotates counter clockwise around the moving
pivot $P_\mathrm{fg}$. The righthand panel shows the polarization
angle as a function of $\lambda^2$, plotted along the lefthand
vertical axis. The total polarization and Stokes $Q$ of the foreground are
plotted along the righthand vertical axis.}
\label{brentjens_fig:slab_delta_polangle}
\end{figure*}

Following \citet{Burn1966}, we make a clear distinction between
Faraday depth ($\phi$) and rotation measure (RM). We define the
Faraday depth of a source as 
\begin{equation}
 \phi(\vec{r}) =
0.81\int_\mathrm{there}^\mathrm{here}n_\mathrm{e} \vec{B}\cdot \
\mathrm{d}\vec{r}\ \mbox{rad}\ \mbox{m}^{-2},
\label{brentjens_eqn:faraday_depth}
\end{equation}
where $n_\mathrm{e}$ is the electron density in $\mbox{cm}^{-3}$,
$\vec{B}$ is the magnetic induction in $\mu$Gauss, and $\
\mathrm{d}\vec{r}$ is an infinitesimal path length in parsecs. A positive
Faraday depth implies a magnetic field pointing towards the observer.
There may exist many different sources of radiation at different
Faraday depths along the same line of sight. These sources may be
either Faraday thin or Faraday thick. A source is Faraday thin if
$\lambda^2\Delta\phi \ll 1$. $\Delta\phi$ denotes
the extent of the source in $\phi$. Faraday thin sources are well
approximated by Dirac $\delta$-functions of $\phi$. A source is
Faraday thick if $\lambda^2\Delta\phi \gg 1$. Faraday thick
sources are extended in $\phi$. They are substantially depolarized at
$\lambda^2$. Remember that whether a source is Faraday thick or Faraday
thin is wavelength dependent.  See
Fig.~\ref{brentjens_fig:slab_delta_polangle} and appendix
\ref{brentjens_sec:example_simulations} for examples.

The rotation measure is commonly defined as the slope of a
polarization angle $\chi$ versus $\lambda^2$ plot:
\begin{equation}
\mbox{RM} = \frac{\mathrm{d}\chi(\lambda^2)}{\mathrm{d}\lambda^2},
\end{equation}
where
\begin{equation}
\chi = \frac{1}{2} \tan^{-1}\frac{U}{Q}.
\end{equation}

\citet{Burn1966} also introduces the complex Faraday dispersion
function $F(\phi)$, which is defined through
\begin{equation}
P(\lambda^2) = \int_{-\infty}^{+\infty}
F(\phi)\mathrm{e}^{2\mathrm{i}\phi\lambda^2}
\ \mathrm{d}\phi.
\label{brentjens_eqn:faraday_dispersion}
\end{equation}
$F(\phi)$ is the complex polarized surface brightness per unit Faraday
depth, and $P(\lambda^2) = p(\lambda^2)I(\lambda^2)$ is the
complex polarized surface brightness. Burn assumes that $F(\phi)$ is
independent of frequency. In section
\ref{brentjens_sec:spectral_dependence} we investigate to what extent
this assumption can be relaxed.

$P$ can be written as
\begin{equation}
P = \|p\|I\mathrm{e}^{2\mathrm{i}\chi},
\label{brentjens_eqn:p_exp}
\end{equation}
or equivalently,
\begin{equation}
P = pI = Q + \mathrm{i} U.
\label{brentjens_eqn:p_qiu}
\end{equation}

Equation (\ref{brentjens_eqn:faraday_dispersion}) is very
similar to a Fourier transform. A fundamental difference is that
$P(\lambda^2)$ only has physical meaning for $\lambda^2 \ge 0$.
Because $P$ cannot be measured at  $\lambda^2 < 0$, equation
(\ref{brentjens_eqn:faraday_dispersion}) is only invertible if one
makes some assumptions on the value of $P$ at $\lambda^2 < 0$
based on its value at $\lambda^2 \ge 0$ \citep{Burn1966}. An example
is that $P$ is Hermitian. This corresponds to assuming that $F(\phi)$
is strictly real. Burn showed that his approach worked by performing the
inversion successfully on the Crab nebula. However, in his derivation
of the inverse, he did not consider the effect of incomplete sampling
of the domain $\lambda^2 > 0$.  In section
\ref{brentjens_sec:derivation} we treat a generalization of equation
(\ref{brentjens_eqn:faraday_dispersion}) that is both invertible and
takes arbitrary sampling of $\lambda^2$ space into account. We refer to
appendix \ref{brentjens_sec:example_simulations} for an example
illustrating the effect of different sampling domains.

If there is only one source along the line of sight, which
in addition has no internal Faraday rotation, and does not suffer from
beam depolarization, then the Faraday depth of that source is equal to
its rotation measure at all wavelengths:
\begin{equation}
\chi(\lambda^2) = \chi_0 + \phi\lambda^2.
\label{brentjens_eqn:polangle}
\end{equation}
In general, however, this is not the case \citep[e.g.][]{Vallee1980}.

A simple example illustrates this. Imagine a classical double
radio galaxy, of which the lobe closest to us is at a Faraday depth of
$\phi_\mathrm{l}\ \mbox{rad}\ \mbox{m}^{-2}$. The lobe itself is
Faraday thin and has an intrinsic polarized flux density of
0.25~Jy~beam$^{-1}$ (positive Stokes $Q$). At low frequencies, there
is usually some polarized Galactic foreground emission between us and
the radio galaxy. The Galactic foreground is modelled as a uniform slab with a constant,
uniform magnetic field. The total integrated polarized surface
brightness of the Galactic foreground at $\lambda=0$ is 1~Jy~beam$^{-1}$
(positive Stokes $Q$). The Faraday dispersion function $F(\phi)$ is a
top hat function:
\begin{equation}
F(\phi) = \left\{
    \begin{array}{ll}
    (2\phi_\mathrm{fg})^{-1} & -\phi_{\mathrm{fg}} \le \phi \le
    +\phi_{\mathrm{fg}}\\
    0 & \mathrm{elsewhere}
    \end{array}
\right. .
\end{equation}

In order to let the foreground emission start at $\phi \ne
0$ we assume a Faraday rotating, but non emitting medium between us
and the Galactic foreground emission. For the sake of simplicity we
assume that the total intensity spectra of both sources are flat. The
complex polarized flux of this configuration is
\begin{eqnarray}
P(\lambda^2) & = & P_\mathrm{fg}(\lambda^2) +
P_\mathrm{l}(\lambda^2)\\
& = &
(2\phi_\mathrm{fg})^{-1}\int_{-\phi_{\mathrm{fg}}}^{+\phi_{\mathrm{fg}}}
\mathrm{e}^{2\mathrm{i}\phi\lambda^2}\ \mathrm{d}\phi +
\frac{1}{4}\mathrm{e}^{2\mathrm{i}\phi_\mathrm{l}\lambda^2}\\
&=& \frac{\sin 2 \phi_{\mathrm{fg}}
\lambda^2}{2\phi_\mathrm{fg}\lambda^2} +
\frac{1}{4}\cos(2\phi_\mathrm{l}\lambda^2) +
\frac{1}{4}\mathrm{i}\sin(2\phi_\mathrm{l}\lambda^2),
\label{brentjens_eqn:slab_polarized_flux}
\end{eqnarray}
where the first term is the contribution of the Galactic foreground
and the last two terms are due to the radio lobe. The first term in
equation (\ref{brentjens_eqn:slab_polarized_flux}) is also called the
Burn depolarization function.  The result for the uniform slab, and
results for several other simple models can be found in
\citet{GardnerWhiteoak1966} and \citet{Burn1966}.
$\ P_\mathrm{fg}(\lambda^2)$ is real because the $F(\phi)$ of
the Galactic foreground emission is symmetric around 0.

Fig.~\ref{brentjens_fig:slab_delta_polangle} plots $\chi$, $\|P\|$,
and $Q_\mathrm{fg}$ for equation
(\ref{brentjens_eqn:slab_polarized_flux}). $Q_\mathrm{fg}$ is the real
part of $P_\mathrm{fg}$. We have taken $\phi_\mathrm{l} = +10\
\mbox{rad}\ \mbox{m}^{-2}$ and $\phi_{\mathrm{fg}} = 2\ \mbox{rad}\
\mbox{m}^{-2}$. At low $\lambda^2$, the foreground dominates over the
lobe, forcing Stokes $Q$ of the sum of the polarizations to be
positive, while $U$ can be both positive and negative. In this regime,
$\chi$ oscillates around zero. However, when the foreground is
significantly depolarized, the lobe starts to dominate the total ($Q$,
$U$) vector. This point is reached somewhere near $\lambda^2 =
0.55$~m$^{2}$. From there on the total ($Q$, $U$) vector runs through
all four quadrants. As the polarized flux of the foreground vanishes,
the total polarization angle approaches more and more a straight line
corresponding to a RM of +10~rad~m$^{-2}$.

\begin{figure}
\resizebox{\hsize}{!}{\includegraphics{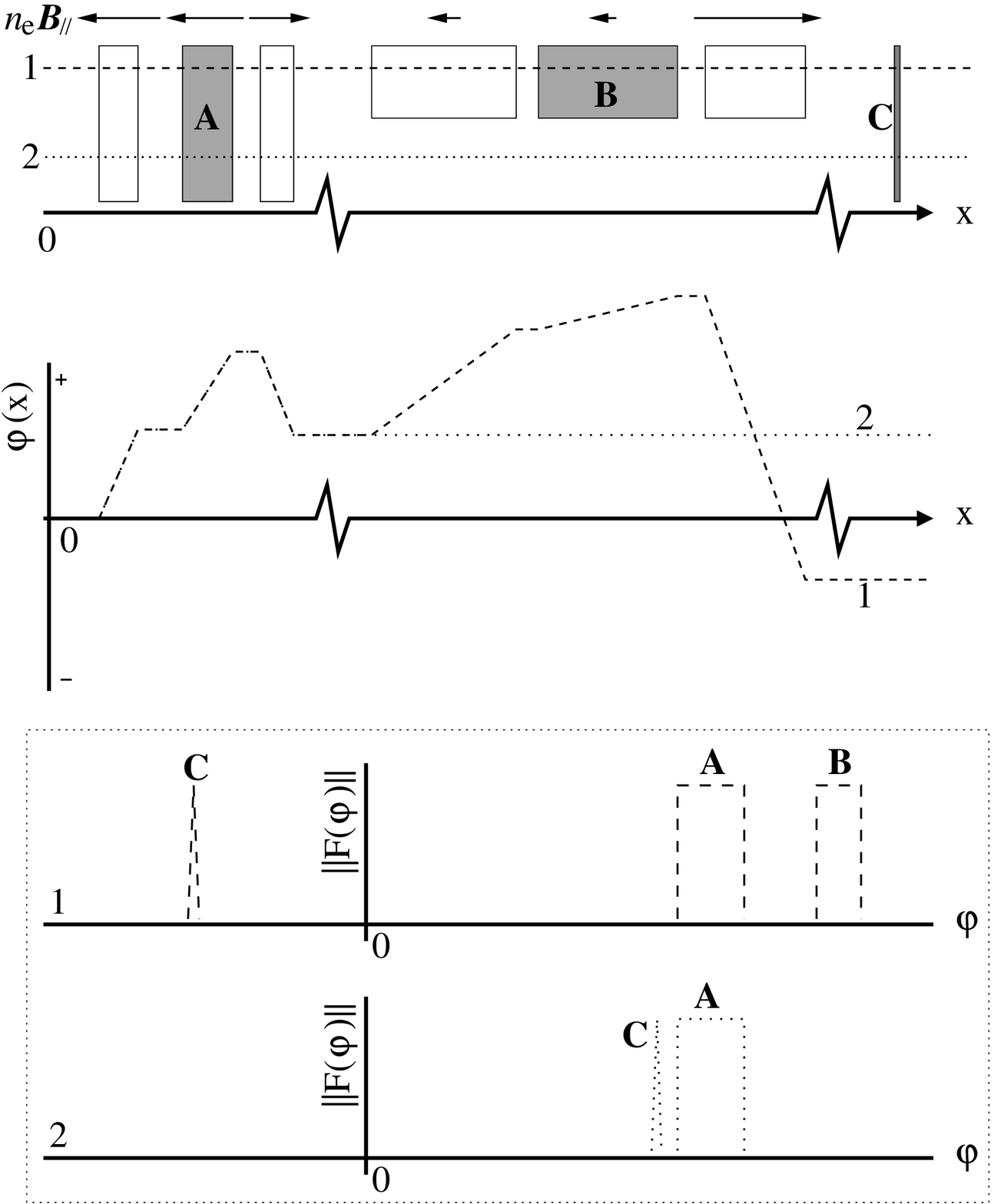}}
\caption{Cartoon sketching the relation between emission,
$n_e\vec{B}_\parallel$, Faraday depth $\phi$, location $x$, and
observed Faraday spectrum. The top panel depicts the physical
situation. The arrows represent $n_\mathrm{e}\vec{B}_\parallel$.
Longer arrows mean larger $\|n_\mathrm{e}\vec{B}_\parallel\|$. The
direction of the arrow indicates the direction of the parallel
component of the magnetic field.  The $x$ coordinate represents
physical distance from the observer. The observer is located at the
far left of the plots. The $x$ axis is severely compressed in two
places. Empty areas have neither emission nor rotation. White blocks
represent areas with only Faraday rotation.  Grey areas with an arrow
have both emission and rotation (area A and B) and grey areas without
an arrow have only emission (area C). There are two lines of sight,
labelled 1 and 2.  Line of sight 1 goes through areas A, B, and C. Line
of sight 2 misses area B as well as the adjacent non emitting Faraday
rotating white boxes. The middle panel plots Faraday depth $\phi$
as a function of physical distance $x$ for both lines of sight. The
bottom panel shows $\|F(\phi)\|$, the observed polarized surface
brightness $\left(\mbox{rad}\ \mbox{m}^{-2}\right)^{-1}$, for both
lines of sight. The peaks in the spectra are labelled with the
associated areas.}
\label{brentjens_fig:depth_versus_x}
\end{figure}

Fig.~\ref{brentjens_fig:depth_versus_x} shows an example of a fairly
complex line of sight. There are three areas with polarized emission
(A, B, and C), of which two (A and B) also have internal Faraday
rotation. The middle panel shows the non-monotonic relation between
Faraday depth and physical depth. Although area B is larger in
physical depth, area A is larger in Faraday depth due to the high
absolute value of $n_\mathrm{e}\vec{B}_\parallel$.

A physical interpretation of this example would be that region A and
its adjacent rotation-only areas reside in our Galaxy, area B and its
neighboring rotation-only areas are a galaxy cluster, and area C
represents a collection of distant polarized sources without any
internal Faraday rotation of their own. Line of sight 1 goes through
the cluster, while line of sight 2 just misses it. This causes C to be
at different Faraday depth in the two lines of sight.

Because of the Fourier nature of both equation
(\ref{brentjens_eqn:faraday_dispersion}) and radio synthesis imaging,
there exist many analogies between the two. Examples are
$uv$ plane sampling versus $\lambda^2$ sampling and synthesized beam
versus RMTF. Therefore  we prefer to call the process of inversion
`Rotation Measure synthesis' (`RM-synthesis' for short).

Similar methods have recently been applied to pulsar observations
\citep{MitraEtAl2003,WeisbergEtAl2004}. \citet{DeBruyn1996RM} applied
the method for the first time to an entire field of view. He also
introduced the concept of a Rotation Measure Transfer Function (RMTF,
see also section \ref{brentjens_sec:derivation} of this work). When
applied to a complete field of view instead of just one line of sight,
the output of a RM-synthesis is a so-called `RM-cube'. The RM-cube
has axes $\alpha$, $\delta$, and $\phi$. It is the Faraday rotation
equivalent of a 21 cm line cube. The application to wide fields
allowed the discovery of widespread, very faint polarized emission
associated with the Perseus cluster \citep{DeBruynBrentjens2005}.

Modern correlator backends, like the ones installed at the WSRT, the
GMRT, and the ATCA and the one to be installed at the EVLA deliver the
visibilities in many (32 to 1024) narrow channels across a wide band
(typically 16 to 160 MHz). The narrow channels move the bandwidth
depolarization limit to much higher rotation measures. The wider bands
yield very high sensitivities if the full bandwidth can be
used. Thanks to these backends RM-synthesis has finally become a
practical, even necessary observing method.

Section \ref{brentjens_sec:derivation} discusses the generally
incomplete sampling of $\lambda^2 > 0$. We formally derive the
RMTF. Section \ref{brentjens_sec:spectral_dependence} treats
modifications to the assumption that $F(\phi)$ is frequency
independent. In section
\ref{brentjens_sec:ambiguities} we treat the relation between the RMTF
and $n\pi$ ambiguities in traditional RM fitting.  Section
\ref{brentjens_sec:qu_only} describes RM-synthesis with Stokes $Q$ or
$U$ only. Section \ref{brentjens_sec:experiment_layout} gives advice
on designing Faraday rotation experiments, taking the findings of this
work into account.  Section \ref{brentjens_sec:conclusions} concludes
this work. Appendix \ref{brentjens_sec:sigma_rm} expands on error
estimation in RM work and appendix
\ref{brentjens_sec:example_simulations} treats an example simulation
illustrating a few important concepts presented in this work.

\section{Derivation}
\label{brentjens_sec:derivation}

\begin{table}
\caption{List of symbols.}
\begin{center}
\begin{tabular}{lp{0.75\hsize}}
\hline
Symbol & Description\\
\hline
$\chi$ & Polarization angle (N through E)\\
 $\chi_0$& Polarization angle at $\lambda = 0$\\
$\nu$  & Frequency\\
 $\delta\nu$ & Channel width in frequency\\
$\nu_\mathrm{c}$ & Central frequency of a channel\\
 $\lambda$ & Wavelength\\
$\lambda_0$ & Wavelength to which all polarization vectors are
derotated\\
$\lambda_\mathrm{c}^2$& Central wavelength squared of a channel\\
$\delta\lambda^2$ & Channel width in wavelength squared\\
$\Delta\lambda^2$ & Total bandwidth in wavelength squared.
$\Delta\lambda^2 = \lambda^2_\mathrm{max} - \lambda^2_\mathrm{min}$\\
$\phi$ & Faraday depth\\
$\delta\phi$ & FWHM of the main peak of the RMTF\\
$\mbox{RM}$ & Rotation measure\\
$W(\lambda^2)$ & Weight function\\
$w_i$ & Weight of the $i$th data point\\
$K$ & One over the integral of $W$ or one over the sum of weights\\
$F(\phi)$ & Faraday dispersion function without spectral dependence\\
$\tilde{F}(\phi)$ & Reconstructed approximation to $F(\phi)$\\
$F(\phi, \lambda^2)$ & General form of the Faraday dispersion function\\
$f(\phi)$ & $F(\phi,\lambda^2)/s(\lambda^2)$\\
$s(\lambda^2)$ & Spectral dependence in $I$, normalized to unity at
$\lambda^2 = \lambda^2_0$\\ 
$\alpha$ & Frequency spectral index\\
$P(\lambda^2)$ & Complex polarized surface brightness\\
$\tilde{P}(\lambda^2)$ & Observed $P$: $W(\lambda^2)P(\lambda^2)$\\  
$p(\lambda^2)$ & Complex polarization fraction
$P(\lambda^2)/I(\lambda^2)$\\
$R(\phi)$ & Rotation Measure Transfer Function (RMTF)\\
$\vec{B}$ & Magnetic induction\\
$\vec{r}$ & Position vector\\
$n_\mathrm{e}$& Thermal electron density\\
$\gamma$ & Spectral index of the relativistic electron energy distribution\\
$\Re z$ & Real part of $z$\\
$\Im z$ & Imaginary part of $z$\\
$\rho$ & Merit function for traditional linear least squares fitting of rotation measures. Defined in equation
(\ref{brentjens_eqn:rho})\\
$\sigma$ & RMS noise in a single channel map\\
$\sigma_\mathrm{Q}$, $\sigma_\mathrm{U}$ & RMS noise in single $Q$ or
$U$ channel maps\\
$\sigma_\mathrm{P}$, $\sigma_\chi$ & Standard error
of $\|P\|$ and $\chi$ in individual channel maps\\
$\sigma_\phi$, $\sigma_{\chi_0}$ & Standard error in Faraday depth
and position angle at $\lambda=0$\\
$\sigma_{\lambda^2}$ & Standard deviation of the distribution of
$\lambda^2$ values that are sampled. This is a measure of the
effective width of the $\lambda^2$ sampling\\
$\delta(x)$ & Dirac delta function\\
\hline
\end{tabular}
\end{center}
\label{brentjens_tab:list_of_symbols}
\end{table}

The goal of this section is to approximate 
$F(\phi)$ by Fourier inverting a generalized version of equation
(\ref{brentjens_eqn:faraday_dispersion}).
Table~\ref{brentjens_tab:list_of_symbols} summarizes the symbols that
are used throughout this paper. We generalize equation
(\ref{brentjens_eqn:faraday_dispersion}) by introducing the weight
function $W(\lambda^2)$. $W(\lambda^2)$ is also called the sampling
function. It is nonzero at all $\lambda^2$ points where measurements
are taken. It is zero elsewhere. Obviously, $W(\lambda^2)=0$ for
$\lambda^2 < 0$  because of the lack of measurements there. The
observed polarized flux density, or surface brightness in the case of
spatially extended sources, is 
\begin{equation}
\label{brentjens_eqn:p_observed}
\tilde{P}(\lambda^2)   =  W(\lambda^2)P(\lambda^2).
\end{equation}
The tilde is used to indicate observed or reconstructed quantities.
Substituting equation (\ref{brentjens_eqn:faraday_dispersion}) gives
\begin{equation}
\label{brentjens_eqn:p_observed_fourier}
\tilde{P}(\lambda^2) = W(\lambda^2)\int_{-\infty}^{+\infty}
F(\phi)\mathrm{e}^{2\mathrm{i}\phi\lambda^2}
\ \mathrm{d}\phi.
\end{equation}

Equation (\ref{brentjens_eqn:faraday_dispersion}) is very similar to
the Fourier transform pair
\begin{eqnarray}
\label{brentjens_eqn:fourier}f(x) & = & \int_{-\infty}^{+\infty}
F(t)\mathrm{e}^{-2\pi\mathrm{i}xt}\ \mathrm{d}t\\
F(t) & = & \int_{-\infty}^{+\infty} f(x)\mathrm{e}^{2\pi\mathrm{i}xt}
\ \mathrm{d}x.
\label{brentjens_eqn:fourier_inverse}
\end{eqnarray}
Substituting $\lambda^2 = \pi u$ in equation
(\ref{brentjens_eqn:p_observed_fourier}) gives
\begin{equation}
\tilde{P}(\pi u) = W(\pi u)\int_{-\infty}^{+\infty}
F(\phi)\mathrm{e}^{2\pi\mathrm{i}\phi u}\ \mathrm{d}\phi.
\label{brentjens_eqn:p_observed_u}
\end{equation}
We define the function
\begin{equation}
R(\phi) = \frac{\int_{-\infty}^{+\infty}W(\pi
u)\mathrm{e}^{-2\pi\mathrm{i}\phi
u}\ \mathrm{d}u}{\int_{-\infty}^{+\infty}W(\pi u)\ \mathrm{d}u},
\end{equation}
which is normalized to unity at $\phi = 0$. The inverse is
\begin{equation}
W(\pi u) =\left(\int_{-\infty}^{+\infty}W(\pi u)\ \mathrm{d}u\ 
\right)\ \int_{-\infty}^{+\infty}R(\phi)\mathrm{e}^{2\pi\mathrm{i}\phi
u}\ \mathrm{d}\phi.
\label{brentjens_eqn:weight_transform}
\end{equation}

Equations (\ref{brentjens_eqn:p_observed_u}) and
(\ref{brentjens_eqn:weight_transform}) are now combined. Application
of  the convolution theorem to the result gives
\begin{equation}
F(\phi)\ast R(\phi) = \frac{\int_{-\infty}^{+\infty}\tilde{P}(\pi
u)\mathrm{e}^{-2\pi\mathrm{i}\phi u}\
\mathrm{d}u}{\int_{-\infty}^{+\infty}W(\pi u)\ \mathrm{d}u},
\label{brentjens_eqn:inversion_t_u}
\end{equation}
where $\ast$ denotes convolution. After back substituting $u =
\lambda^2/\pi$, we obtain 
\begin{eqnarray}
\tilde{F}(\phi) = F(\phi)\ast R(\phi) & =
& K\int_{-\infty}^{+\infty}\tilde{P}(\lambda^2)\mathrm{e}^{-2\mathrm{i}\phi
\lambda^2}\ \mathrm{d}\lambda^2\label{brentjens_eqn:inversion_0}\\
R(\phi) & =& K\int_{-\infty}^{+\infty}W(\lambda^2) \mathrm{e}^{-2
\mathrm{i} \phi \lambda^2}\ \mathrm{d}\lambda^2
\label{brentjens_eqn:rmtf_0}\\
K & = & \left(\int_{-\infty}^{+\infty}W(\lambda^2)\ \mathrm{d}
\lambda^2\right)^{-1}.
\label{brentjens_eqn:K}
\end{eqnarray}
$\tilde{F}(\phi)$ is an approximate reconstruction of $F(\phi)$. More
precisely, it is $F(\phi)$ convolved with $R(\phi)$ after Fourier
filtering by the weight function $W(\lambda^2)$. The
quality of the reconstruction depends mainly on the weight function
$W(\lambda^2)$. A more complete coverage of $\lambda^2$ space improves
the reconstruction. Fewer holes in the $\lambda^2$ sampling reduce the
side lobes of $R(\phi)$, while covering a larger range of $\lambda^2$
increases the resolution in $\phi$ space. We return to these
statements in the following sections. We call $R(\phi)$ the
rotation measure transfer function (RMTF).

\begin{figure}[t]
\resizebox{\hsize}{!}{\includegraphics{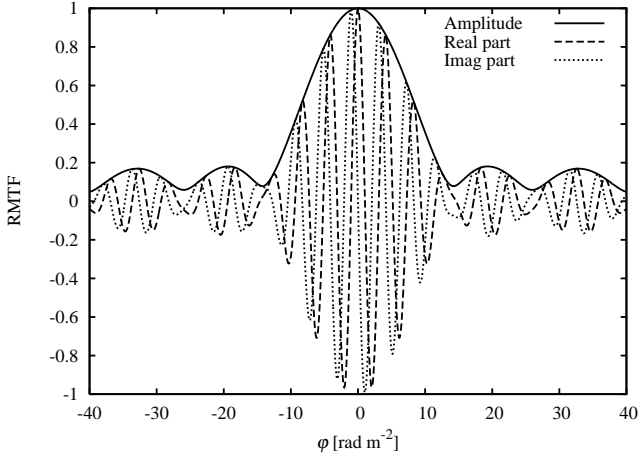}}
\caption{RMTF of a 92 cm dataset taken with the Westerbork Synthesis
Radio Telescope. There are 126 usable channels in the dataset. All
($Q$,$U$) vectors have been derotated to $\lambda^2_0 = 0$. Note
the rapid rotation of the RMTF, making it difficult to measure
accurate polarization angles in a sampled RM cube.}
\label{brentjens_fig:rmtf_0}
\end{figure}

The above set of equations is not yet our final result.
Fig.~\ref{brentjens_fig:rmtf_0} displays the rotation measure transfer
function corresponding to the $\lambda^2$ sampling of our Perseus data
set \citep{DeBruynBrentjens2005}. It only shows a small part of the
RMTF close to the peak. The response function displays a rapid
rotation of the (real, imaginary) vector. Because one usually samples
$\phi$ space at finite intervals, this rotation makes it very
difficult to correctly estimate the polarization angle at or near the
maximum of $\|F(\phi)\|$. If the Faraday depth of a frame is only a tenth
of the width of the RMTF away from the actual Faraday depth of the
source, the (real, imaginary) vector may already be rotated by
several tens of degrees.

Equations (\ref{brentjens_eqn:inversion_0}) and
(\ref{brentjens_eqn:rmtf_0}) correspond to derotating all polarization
vectors back to their position at $\lambda = 0$. At first this appears
sensible, because the polarization vector at $\lambda = 0$ is directly
related to the electric field vector in the plane of the
sky without any Faraday rotation. Nevertheless no information is lost by
derotating to some other common $\lambda^2_0 \ne 0$. 

The more general versions of equations
(\ref{brentjens_eqn:inversion_0}) and (\ref{brentjens_eqn:rmtf_0}) are
\begin{eqnarray}
\tilde{F}(\phi) & = & K \int_{-\infty}^{+\infty}\tilde{P}
(\lambda^2)\mathrm{e}^{-2\mathrm{i}\phi (\lambda^2-\lambda_0^2)}\ 
\mathrm{d}\lambda^2 \label{brentjens_eqn:inversion}\\
R(\phi) & =& K \int_{-\infty}^{+\infty}W(\lambda^2) \mathrm{e}^{-2
\mathrm{i} \phi (\lambda^2-\lambda_0^2)}\
\mathrm{d}\lambda^2\label{brentjens_eqn:rmtf}.
\end{eqnarray}

This is effectively an application of the shift theorem of
Fourier theory. Because the shift theorem only affects the argument,
and not the absolute value of the resulting complex function, nothing
changes in the amplitude of the RMTF.  Equations
(\ref{brentjens_eqn:inversion}) and
(\ref{brentjens_eqn:p_observed_fourier}) form a Fourier pair that
enables us to transform polarization information from $\lambda^2$
space to $\phi$ space and back. The function $R(\phi)$ is
our final form of the rotation measure transfer function (RMTF). It is
a complex valued function. The real part corresponds to the response
of the transform parallel to the $(Q,U)$ vector at $\lambda =
\lambda_0$ and the imaginary part corresponds to the response
orthogonal to it.  Assume that one has a Faraday thin source
at Faraday depth $\phi_0$, of which the polarization angle is
45$\degr$ (all polarized emission is in positive $U$) at $\lambda^2 =
\lambda^2_0$. Whenever $R(\phi-\phi_0)$ is real and positive,
$\tilde{F}(\phi)$ would be imaginary (all polarized flux in positive
Stokes $U$). If, however $R(\phi)$ is positive imaginary,
$\tilde{F}(\phi)$ would be real and negative (all polarized flux in
negative Stokes $Q$. This is important if $\phi \ne \phi_0$, that is,
when the Faraday depth of the source does not match the Faraday depth
that was chosen for evaluation of
$\tilde{F}(\phi)$.

The simplest way to see this is to consider the case when $F(\phi) =
P(\lambda_0^2)\delta(\phi -\phi_\mathrm{source})$. This changes the
convolution in equation (\ref{brentjens_eqn:inversion}) into a
multiplication. Hence the result of the righthand side of equation
(\ref{brentjens_eqn:inversion}) can be written as
\begin{equation}
\tilde{F}(\phi) = P(\lambda_0^2) R(\phi -\phi_\mathrm{source}).
\end{equation}
If $R$ is imaginary, it can be written as
$\|R\|\mathrm{e}^{\pm\mathrm{i}\pi/2}$. Multiplication of a
complex number with $\mathrm{e}^{\pm\mathrm{i}\pi/2}$
corresponds to a rotation in the complex plane of $\pm\pi/2$, hence
the apparent polarization angle would have rotated by 45$\degr$
relative to the actual polarization angle at $\lambda^2 =
\lambda^2_0$.

\begin{figure}[t]
\resizebox{\hsize}{!}{\includegraphics{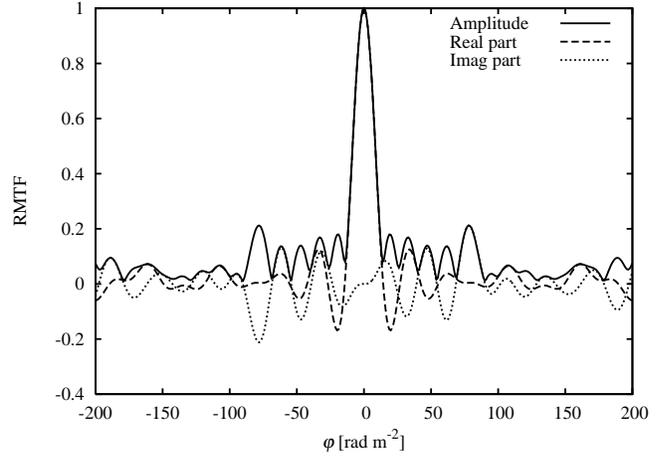}}
\caption{RMTF of the same dataset as described in
Fig.~\ref{brentjens_fig:rmtf_0}. This time, however, all $\mathbf{P}$
vectors have been derotated to the average $\lambda^2$. It is seen
that the imaginary part remains almost constant within the
central peak of the RMTF.}
\label{brentjens_fig:rmtf_avg_zoom}
\end{figure}

Ideally, the response in the \emph{entire} main peak of the RMTF
should be parallel to the actual polarization vector at
$\lambda_0$. The best way of achieving that is keeping the orthogonal
response as close to zero as possible. We set the derivative of the
imaginary part at $\phi = 0$ to zero:
\begin{eqnarray}
0 & = & \left.\frac{\partial\Im{R(\phi)}}{\partial\phi}\right|_{\phi=0}\\
0 & = &\left.-K
\frac{\partial}{\partial\phi}\int_{-\infty}^{+\infty}W(\lambda^2) \sin
2\phi(\lambda^2 - \lambda_0^2)\ \mathrm{d}\lambda^2\right|_{\phi=0} \\
0 & = & \left.-K\int_{-\infty}^{+\infty} W(\lambda^2) 2(\lambda^2 -
\lambda_0^2)\cos 2\phi(\lambda - \lambda_0^2)\ \mathrm{d}\lambda^2\right|_{\phi=0}\\
0 & = &\int_{-\infty}^{+\infty} W(\lambda^2) (\lambda^2 - \lambda_0^2)\ \mathrm{d}\lambda^2\\
\lambda_0^2 & = & \frac{\int_{-\infty}^{+\infty} W(\lambda^2) \lambda^2\
\mathrm{d}\lambda^2}{\int_{-\infty}^\infty W(\lambda^2)\
\mathrm{d}\lambda^2},
\label{brentjens_eqn:average_lambda2}
\end{eqnarray}
hence $\lambda_0^2$ should be made equal to the weighted average
of the observed $\lambda^2$.

A drawback of having $\lambda_0^2 \ne 0$ is that the polarization
angle that one derives still needs to be transformed to a polarization
angle at $\lambda^2 = 0$, if one wants information on the orientation
of the electric field direction in the source. In case of a
high S/N ratio, this is very easy: 
\begin{equation}
\chi_0 = \chi(\lambda_0^2) - \phi\lambda_0^2.
\end{equation}
However, if the signal to noise ratio is low, the uncertainty in
$\phi$ usually prevents accurate derotation to $\lambda^2 = 0$. The
advantage of derotating to the weighted average $\lambda^2$ is that
one can still properly analyze spatial coherence  of polarization
angles in a spatially extended source at a certain Faraday depth.

Fig.~\ref{brentjens_fig:rmtf_avg_zoom} shows the same RMTF as
Fig.~\ref{brentjens_fig:rmtf_0}, except that $\lambda_0^2$ is set to
the weighted average $\lambda^2$. The improvement with respect to the
orthogonal response is evident. The response function is almost
completely real between the first minima. The only drawback is that
one cannot convert the observed polarization angle at $\lambda_0$ to a
$\vec{E}$ vector in a straightforward way. In order to accomplish
reliable derotation to $\lambda = 0$, one needs a sufficiently high
S/N ratio to determine the Faraday depth with an accuracy well within
the full width at half maximum (FWHM) of the RMTF. This is
not a problem for bright sources that are already detected in
individual channels, but for faint emission that is only detectable
after RM-synthesis, one cannot usually do this. These signal to noise
statements are quantified in section \ref{brentjens_sec:ambiguities}
and appendix \ref{brentjens_sec:sigma_rm}.

In most correlators, all channels have equal bandwidth $\delta\nu$, centred
around $\nu_\mathrm{c}$, the central frequency of the channel. Our
prime coordinate is $\lambda^2$, not $\nu$. If we assume a top hat
channel bandpass, we have for every channel:
\begin{equation}
\lambda^2_\mathrm{c} \approx \frac{c^2}{\nu_\mathrm{c}^2}\left(
1+\frac{3}{4} \left(\frac{\delta\nu}{\nu_\mathrm{c}}\right)^2 \right)
\label{brentjens_eqn:lambda2_c}
\end{equation}
and
\begin{equation}
\delta\lambda^2 \approx \frac{2c^2\delta\nu}{\nu_\mathrm{c}^3}
\left(1+ \frac{1}{2}\left(\frac{\delta\nu}{\nu_\mathrm{c}}\right)^2\right).
\label{brentjens_eqn:delta_lambda2}
\end{equation}

Of course the channel bandpass is usually not a top hat
function, but rather a sinc if no taper was applied before the
time-to-frequency transform, or something in between if other tapers
like Hanning or Kaiser-Bessel are used. These differences are hardly
important as long as $\delta\nu/\nu_\mathrm{c} \ll 1$, which is usually
the case.

If $\phi\delta\lambda^2 \ll 1$ for all channels, we may approximate
the integrals in equations (\ref{brentjens_eqn:inversion}) and
(\ref{brentjens_eqn:rmtf}) by sums:
\begin{eqnarray}
\tilde{F}(\phi) & \approx
&K \sum_{i=1}^{N}\tilde{P}_i\mathrm{e}^{-2\mathrm{i}\phi
(\lambda^2_i-\lambda_0^2)} \label{brentjens_eqn:inversion_sum}\\
\nonumber\hspace{1cm}\\
R(\phi) & \approx &K \sum_{i=1}^{N}w_i\mathrm{e}^{-2
\mathrm{i} \phi (\lambda^2_i-\lambda_0^2)}
\label{brentjens_eqn:rmtf_sum}\\
K & = &\left(\sum_{i=1}^{N} w_i\right)^{-1}.
\label{brentjens_eqn:sum_of_weights}
\end{eqnarray}
In these equations, $\lambda^2_i$ is $\lambda^2_\mathrm{c}$ of channel
$i$, $\tilde{P}_i = \tilde{P}(\lambda^2_i) = w_iP(\lambda_i^2)$, $w_i
= W(\lambda^2_i)$, and $K$ has become the sum of all weights.
We have implemented equations
(\ref{brentjens_eqn:inversion_sum}), (\ref{brentjens_eqn:rmtf_sum}),
and (\ref{brentjens_eqn:sum_of_weights}) in our RM-synthesis
software.

\section{Spectral dependence}
\label{brentjens_sec:spectral_dependence}

In this section we investigate the effect of the emission spectrum of
a source on the method. We start with the most general case of an
arbitrary spectrum at each Faraday depth. We substitute
\begin{equation}
F(\phi) = F(\phi, \lambda^2)
\end{equation}
in equation (\ref{brentjens_eqn:p_observed_fourier})
\begin{equation}
\tilde{P}(\lambda^2) = W(\lambda^2)\int_{-\infty}^{+\infty}
F(\phi, \lambda^2)\mathrm{e}^{2\mathrm{i}\phi\lambda^2}
\ \mathrm{d}\phi.
\end{equation}
In general, this equation is not invertible, except in cases
such as:
\begin{itemize}
\item $F(\phi, \lambda^2)$ can be written as \modified{a product of
independent functions, i.e. $F(\phi, \lambda^2) =
f(\phi)s(\lambda^2)$};
\item $F(\phi, \lambda^2) =  f(\phi)\delta(\phi-\phi_0)s(\lambda^2)$;
\item the spectrum is a power law with $\alpha \propto \phi$.
\end{itemize}
$s(\lambda^2)$ is the spectral dependence.
\begin{equation}
s(\lambda^2) = \frac{I(\lambda^2)}{I(\lambda_0^2)}
\label{brentjens_eqn:spectral_dependence}
\end{equation}

The third case, although invertible, is
highly non-physical. The inversion of the first case is trivial:
\begin{eqnarray}
F(\phi, \lambda^2) & = & f(\phi)s(\lambda^2)\label{brentjens_eqn:separable}\\
\tilde{P}(\lambda^2) & =& W(\lambda^2)\int_{-\infty}^{+\infty}
f(\phi)s(\lambda^2)\mathrm{e}^{2\mathrm{i}\phi\lambda^2}
\ \mathrm{d}\phi\\
\frac{\tilde{P}(\lambda^2)}{s(\lambda^2)} & =&
\tilde{p}(\lambda^2)I(\lambda_0^2) = W(\lambda^2)
\int_{-\infty}^{+\infty} f(\phi)\mathrm{e}^{2\mathrm{i}\phi\lambda^2}
\ \mathrm{d}\phi.\label{brentjens_eqn:polarization_spectrum}
\end{eqnarray}
This is equivalent to equation
(\ref{brentjens_eqn:p_observed_fourier}) divided by $s(\lambda^2)$,
and we have already shown that equation
(\ref{brentjens_eqn:p_observed_fourier}) is invertible.

The second case is a specialization of the first case. Equation
(\ref{brentjens_eqn:polarization_spectrum}) reduces to equation
(\ref{brentjens_eqn:p_observed_fourier}) in case of a flat
spectrum. The approximate Faraday dispersion function compensated for
a non-flat spectrum is given by:
\begin{equation}
\tilde{f}(\phi) = K  \int_{-\infty}^{+\infty}{
W(\lambda^2)\frac{P(\lambda^2)}{s(\lambda^2)}
\mathrm{e}^{-2\mathrm{i}\phi(\lambda^2 - \lambda_0^2)}\
\mathrm{d}\lambda},
\end{equation}
or, if $W(\lambda^2)$ can be approximated by a sum of Dirac $\delta$
functions: 
\begin{equation}
\tilde{f}(\phi)  \approx
K \sum_{i=1}^{N}\frac{\tilde{P}_i}{s_i}\mathrm{e}^{-2\mathrm{i}\phi
(\lambda^2_i-\lambda_0^2)}, \label{brentjens_eqn:inversion_sum_spectrum}
\end{equation}
where $s_i = s(\lambda^2_i)$ and $\tilde{f}(\phi) = f(\phi)\ast R(\phi)$.


\begin{figure}[t]
\resizebox{\hsize}{!}{\includegraphics{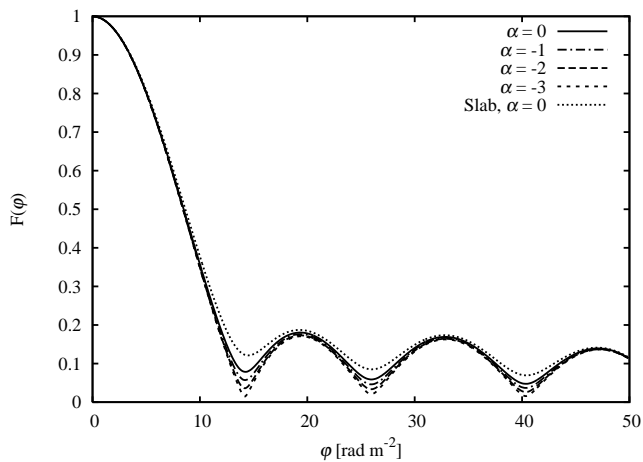}}
\caption{Absolute value of the approximated Faraday dispersion
function of several Faraday thin sources with different spectral
indices. The $\lambda^2$ grid is the same previously mentioned 126
channel dataset. The overall bandwidth $\Delta\nu/\nu = 17\%$.
Steeper spectra give deeper nulls. For comparison the normalized
approximated Faraday dispersion function of a Faraday thick slab was
included.}
\label{brentjens_fig:rmtf_spectrum}
\end{figure}

Equation (\ref{brentjens_eqn:separable}) applies only in some very
specific scenarios. It holds for example in optically thin
synchrotron-emitting and Faraday-rotating clouds that have the same
relativistic electron energy distribution throughout the
cloud. It also holds if multiple optically thin clouds along the line
of sight happen to have the same spectral dependence. Optically thin
synchrotron radiation has a spectrum that is proportional to
$\nu^\alpha$ over a large range of frequencies
\citep{ConwayKellermannLong1963}. For most sources,
$\alpha$ is in the range $\langle -1.5, -0.5\rangle$. In extreme cases
the spectral index of optically thin emission can approach 0
(e.g. the Crab nebula) or -3 (for halo or relic sources in galaxy
clusters). 

In general, spectral indices vary across a map. One can of course
easily correct for the spectra of sources that are reliably detected
in individual channel maps. This is impossible for sources that are
much fainter and only show up after averaging the full band. For
those objects it makes sense to estimate some ``average'' spectral
index and apply that to the entire map.

What is the effect of using the wrong spectral index in correcting for
the spectrum of a single source along the line of sight? The
contributions of multiple sources along the line of sight is simply
the sum of their individual responses. Because
the spectrum is an amplitude only effect, it has no influence on the
location of the maximum of the Faraday dispersion function of the
source.  Therefore its derived Faraday depth is unaffected. It does
distort the RMTF associated with the source at points away from the
main peak.  This complicates deconvolution algorithms slightly.

Fig.~\ref{brentjens_fig:rmtf_spectrum} gives the Faraday
dispersion functions of Faraday thin model sources with spectral
indices $-3$ to 0. It is seen that the largest effect
occurs close to the nulls of the RMTF. The difference
between $\alpha = -3$ and $\alpha = 0$ is small over the 17\% total
frequency bandwidth in the simulation.  It will not be noticeable if
the emission has such low S/N that it is invisible in individual
channels. For comparison, the normalized $\tilde{F}(\phi)$  of a
Faraday thick uniform slab model is included. The slab emits at $-1
\le \phi \le +1\ \mbox{rad}\ \mbox{m}^{-2}$. It is seen that the
effect of even a tiny amount of $\phi$ structure in the source is
much larger than the effect of changing the spectral index by $\pm1$.

The general case of an arbitrary spectral dependence at
multiple Faraday depths is not invertible. One can only recover the
Faraday dispersion function if the spectral dependence is the same at
all Faraday depths along the line of sight. One should then divide the
observed polarization by the spectral dependence in $I$.
Fig.~\ref{brentjens_fig:rmtf_spectrum} shows that if the spectral
index is estimated with an absolute uncertainty less than
$1$, the maximum absolute error of the estimated flux density at a
certain Faraday depth is less than 2--5\% of the brightest emission
along the line of sight. This accuracy is easily exceeded for sources
that are visible in total intensity. Sources that have not been
detected in total intensity should generally be assigned a
spectral index of $-1$. This worked very well in
our observations of the Perseus cluster, where we see large, faint
polarized features that have no detectable counterpart in total
intensity \citep{DeBruynBrentjens2005}.

\begin{figure}
\centering
\resizebox{\hsize}{!}{\includegraphics{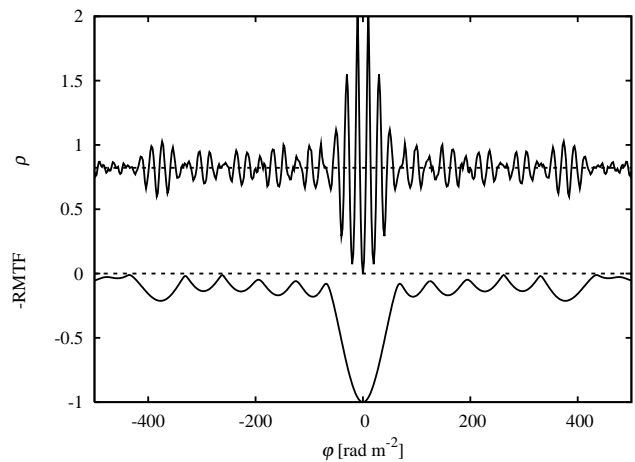}}
\caption{Comparison of merit function $\rho$ (top) and $-\|R\|$
(bottom). The dotted line through $\rho$ is at $y = \pi^2/12$.}
\label{brentjens_fig:rmtf_126pts_zoom}
\end{figure}

\section{$n\pi$ ambiguities and the RMTF}
\label{brentjens_sec:ambiguities}

\begin{figure*}
\centering
\begin{minipage}{0.495\textwidth}
\includegraphics[width=\textwidth]{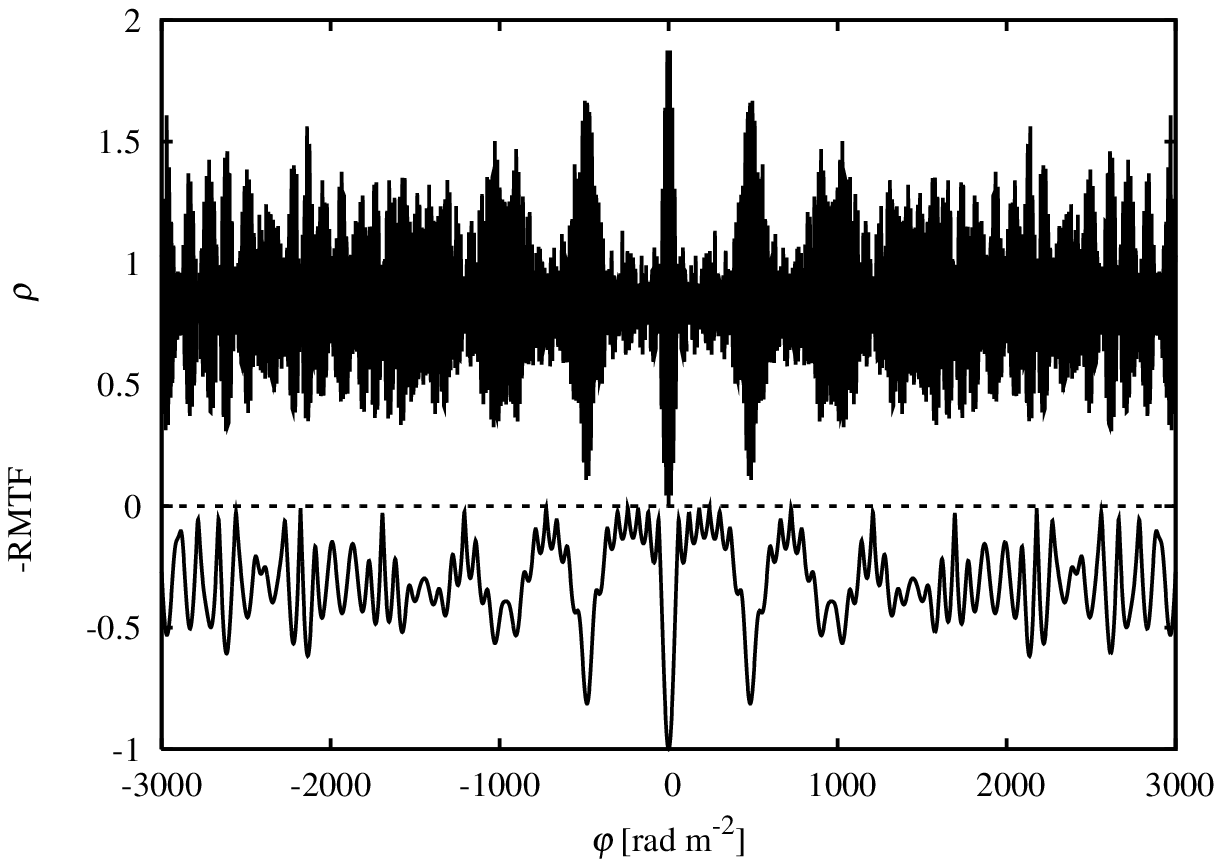}
\end{minipage}
\begin{minipage}{0.495\textwidth}
\includegraphics[width=\textwidth]{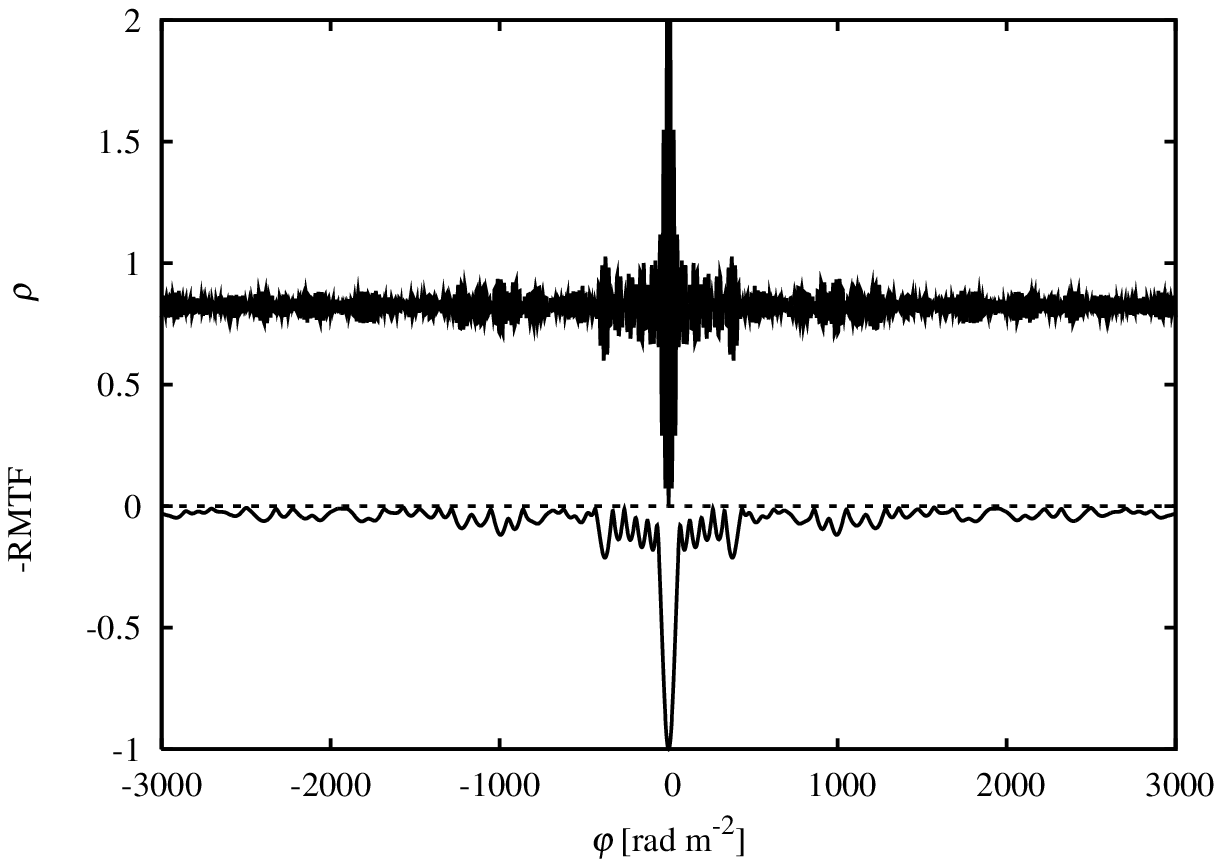}
\end{minipage}
\caption{Both plots show, from top to bottom, merit function $\rho$
and $-\|R(\phi)\|$. The lefthand panel uses eight sample points,
equally spaced in frequency. The frequency sampling of the righthand
panel is identical to the sampling used in
Fig.~\ref{brentjens_fig:rmtf_126pts_zoom}. The sampling in the
lefthand panel has the same extent in $\lambda^2$-space as the
sampling of the righthand panel.}
\label{brentjens_fig:rmtf_126pts}
\end{figure*}

\begin{figure*}
\centering
\begin{minipage}{0.495\textwidth}
\includegraphics[width=\textwidth]{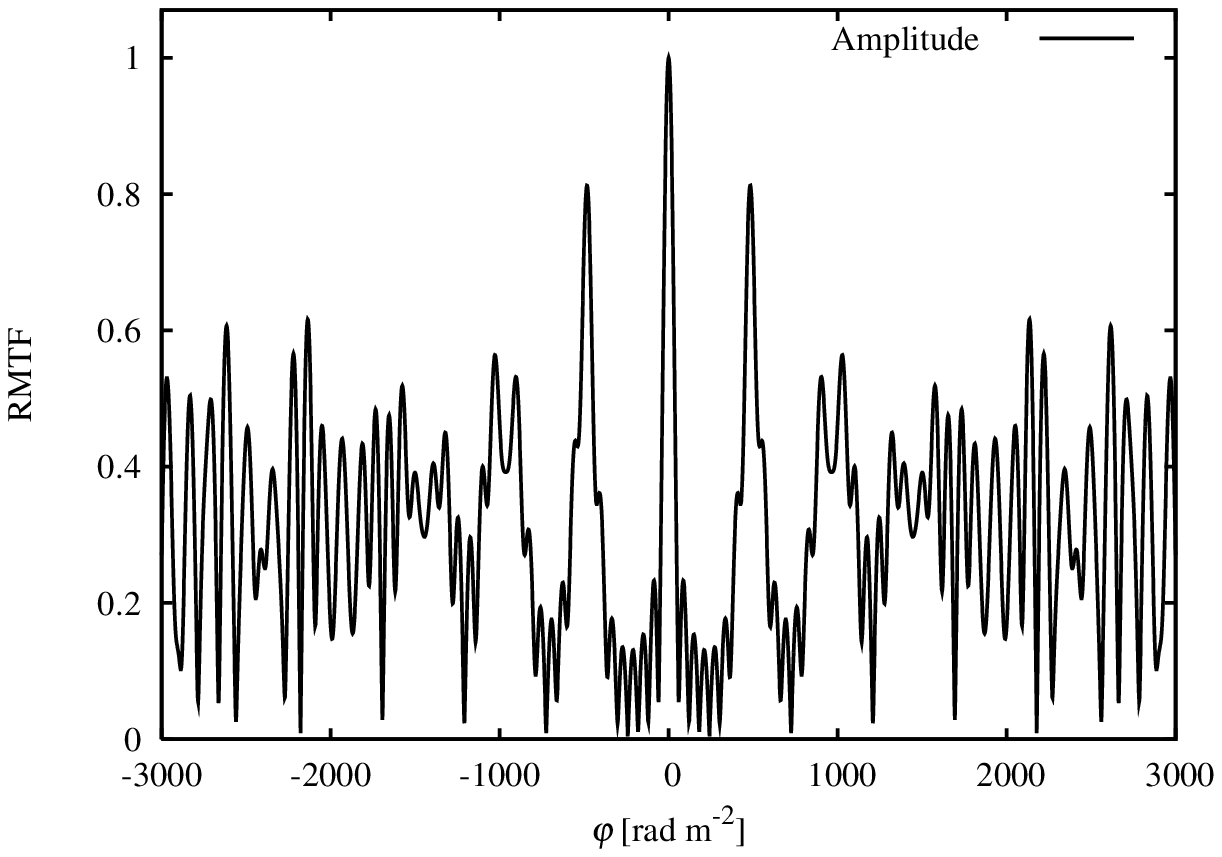}
\end{minipage}
\begin{minipage}{0.495\textwidth}
\includegraphics[width=\textwidth]{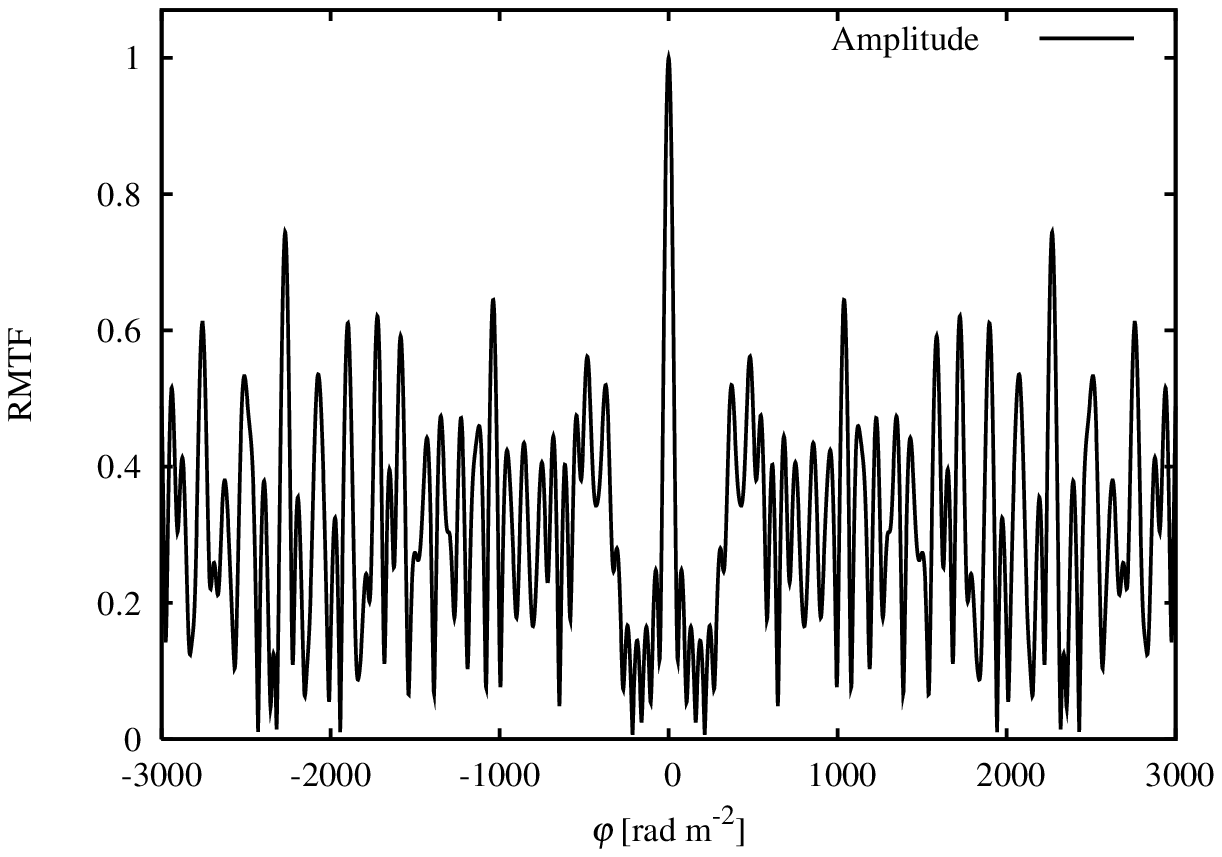}
\end{minipage}
\caption{This plot shows the effect of tweaking the exact frequencies
of eight sampling points. The lefthand panel shows the same RMTF as
the lefthand panel of Fig.~\ref{brentjens_fig:rmtf_126pts}. In the
righthand plot, however, we stretched the frequency intervals such
that low frequency intervals are wider than high frequency
intervals. This eliminates the resonances from the lefthand plot.}
\label{brentjens_fig:rmtf_8pts}
\end{figure*}

The traditional way to compute the rotation measure of a source is to
measure its polarization angle at several wavelengths and determine
the slope of a straight line through the polarization angle as a
function of $\lambda^2$. This method suffers from so-called $n\pi$
ambiguity problems. If only a few data points are available, there may
exist multiple RM solutions that are equally good, but have the
polarization angle of the data points wrapped around one or more
turns. Complicated methods have been devised to attempt to circumvent
these problems, some of which are quite successful. An example is the
``Pacerman'' method \citep{PacmanI,PacmanII}, which operates on images,
and does a good job in finding and correcting $n\pi$ ambiguities using
spatial continuity arguments.

In this section we show that the RMTF is an excellent indicator of
possible $n\pi$ ambiguity problems. By analyzing the RMTF, one can
take measures to minimize or even eradicate any potential $n\pi$
problems in the experiment in advance. We also show that using
\emph{only} RM-synthesis to determine Faraday depths is as
accurate as traditional $\lambda^2$ fitting, but has the added value
of straightforward $n\pi$ ambiguity problem detection.

We first consider traditional $\lambda^2$ fitting. This is done by
linear least squares minimization of a merit function $\rho$. If the
estimated errors of all points are equal, then the merit function 
looks like equation (\ref{brentjens_eqn:rho})
\begin{eqnarray}
z_i &=& \left\|\chi_\mathrm{M}(\mbox{RM}, \chi_0,\lambda^2_i) -
\chi_i\right\|\\
l_i & = & \left\{\begin{array}{lcr}
z_i & &z_i < \frac{1}{2}\pi\\
z_i-\pi & &z_i \ge \frac{1}{2}\pi
\end{array}\right.\label{brentjens_eqn:difference_mod_halfpi}\\
\rho & = & N^{-1} \sum_{i=1}^{N} l_i^2
\label{brentjens_eqn:rho}
\end{eqnarray}
$\chi_\mathrm{M}$ is a polarization angle computed by a model
(for example equation (\ref{brentjens_eqn:polangle})).
$\chi_i$ is the observed polarization angle of the $i$th data point.
Both $\chi_\mathrm{M}$ and $\chi_i$ are modulo $\pi$. Equation
(\ref{brentjens_eqn:difference_mod_halfpi}) ensures that $l_i \in
\left[-\frac{1}{2}\pi, \frac{1}{2}\pi\right\rangle$. For the sake of
simplicity, we assume that we want to fit a source with a constant
polarization angle of 0$\degr$. In that case equation
(\ref{brentjens_eqn:rho}) reduces to
\begin{equation}
\rho = N^{-1} \sum_{i=1}^{N} \left(\left[\mbox{RM}\lambda_i^2\right]
\ \mbox{mod}\ \frac{\pi}{2}\right)^2.
\label{brentjens_eqn:rho_simplified}
\end{equation}
If the fit to the (noiseless) simulated data is perfect, $\rho =
0$. When comparing equation (\ref{brentjens_eqn:rho_simplified}) and
equation (\ref{brentjens_eqn:rmtf_sum}), it is seen that $\rho = 0$ if
and only if $\|R(\phi-\phi_0)\| = 1$. $\phi_0$ denotes the Faraday
depth of the source. In this simple case $\phi =
\mbox{RM}$. If $\|R(\phi-\phi_0)\| < 1$, then $\rho > 0$. 

If the model RM is sufficiently different from the actual RM, one
expects the errors $l_i$ to be approximately uniformly distributed in
the range $\left[-\frac{1}{2}\pi,
\frac{1}{2}\pi\right\rangle$. Because the square is taken, this is
equivalent to a uniform distribution in the range $\left[0,
\frac{1}{2}\pi\right]$. The average value of $l_i^2$ is then
given by
\begin{equation}
\langle l_i^2\rangle = \frac{2}{\pi} \int_0^{\frac{1}{2}\pi} x^2
\mathrm{d}x = \frac{\pi^2}{12}.
\end{equation}

The top graph of Fig.~\ref{brentjens_fig:rmtf_126pts_zoom} plots
$\rho$. The pattern of sample points is that of the 126 points used in
Fig.~\ref{brentjens_fig:rmtf_avg_zoom}. In this case, the width of the
pattern is scaled to $\Delta\lambda^2 = 0.0459\ \mbox{rad}\
\mbox{m}^{-2}$. The bottom graph is $-\|R(\phi-\phi_0)\|$. The dashed
lines are at $y = 0$ and $y = \pi^2/12$. It is seen that
$\rho$ is indeed zero when $\|R\| = 1$. Also, the average value at
high RM is indeed equal to $\pi^2/12$, especially when $\|R\|$
is close to zero. This implies that when $\|R\|$ is close to zero,
$\phi\lambda^2$ is distributed uniformly between $-\frac{1}{2}\pi$ and
$+\frac{1}{2}\pi$. Of course $\|R\|$ can be zero for other reasons,
but due to the large number of points and the dense filling of
$\Delta\lambda^2$, that is rather unlikely.

An interesting aspect seen in
Fig.~\ref{brentjens_fig:rmtf_126pts_zoom} is that the envelope of
$\rho-\pi^2/12$ looks like $-\|R\|$ when $\rho < \pi^2/12$. Deep
minima of $\rho$ are associated with high peaks in the
RMTF. In fact, they appear to be approximately proportional to
$\|R\|$. These deep minima are closely related to so-called
$n\pi$ ambiguities in traditional RM measurements. They
are points that fit the data (almost) equally well as the "true"
solution.

The similarity between the envelope of $\rho$ and $-\|R\|$
is better demonstrated in Fig.~\ref{brentjens_fig:rmtf_126pts}.
It shows both $\rho$ and $-\|R\|$ over a large range in
$\phi$. The lefthand panel displays $\rho$ and $-\|R\|$ for 8 points,
equally spaced in frequency.  To facilitate comparison, the total
width of the pattern, $\Delta \lambda^2$, has been scaled to match the
width of the $\lambda^2$ sampling of
Fig.~\ref{brentjens_fig:rmtf_126pts_zoom}.  The righthand panel of
Fig.~\ref{brentjens_fig:rmtf_126pts} shows $\rho$ and $-\|R\|$ based
on the same input data as Fig.~\ref{brentjens_fig:rmtf_126pts_zoom}.
It is obvious that the RMTF of a 126 point sampling has much lower
side lobes than an 8-point sampling.  $n\pi$ ambiguities are
completely eliminated.

The lefthand panel of Fig.~\ref{brentjens_fig:rmtf_8pts} shows the
same RMTF as the lefthand panel of
Fig.~\ref{brentjens_fig:rmtf_126pts}. The two resonances to the left
and right are due to the near-regularity of the sampling
points in $\lambda^2$ space. If the frequency intervals at the lower
frequencies are stretched more than at the intervals at
higher frequencies, for example by making them decrease linearly with
increasing frequency, one can make the pattern in $\lambda^2$ space
less regular. The result is shown in the righthand panel of
Fig.~\ref{brentjens_fig:rmtf_8pts}.  The resonances are now lower and
wider. If one requires the highest side lobes to be at least 5$\sigma$
lower than unity, then a total S/N of 20 (7 per channel) is sufficient
to prevent $n\pi$ ambiguities outside the main peak of the RMTF. Using
the same requirement, in case of the 126 points, a S/N of 6 in total
(0.6 per channel) is enough to prevent $n\pi$ ambiguities outside the
main peak of the RMTF.

$n\pi$ ambiguities are conceptually closely related to the grating
response of a regularly spaced interferometer like the WSRT. When an
interferometer only has baselines that are a multiple of some fixed
distance, then its instantaneous synthesized beam is a collection of
parallel fan beams. Each fan beam has the same peak amplitude.
Therefore, without any further constraints it is impossible to
determine in which fan beam the source is actually located. The same
holds true for $\lambda^2$ sampling. If one only samples $\lambda^2$
space at regular intervals, there exist multiple solutions that fit
the data equally well. These solutions correspond to peaks of unit
amplitude in the RMTF: grating responses.

Fig.~\ref{brentjens_fig:rmtf_126pts_zoom} also shows that there
are multiple minima of $\rho$ within the main lobe of the RMTF. These
indicate uncertainties in the RM smaller than the width of the main
peak. We shall now investigate the uncertainty in RM within the main
peak of $\|R\|$.

\begin{figure}
\centering
\resizebox{\hsize}{!}{\includegraphics{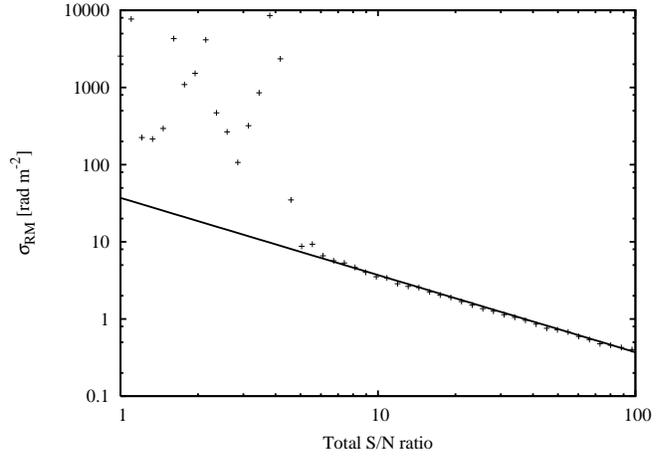}}
\caption{Comparison between the standard error in RM obtained by
traditional line fitting (line) to simulated RM-synthesis experiments
where a parabola was fit to the main peak of the Faraday dispersion
function (dots). The 126 $\lambda^2$ points used for this figure are
the same as the ones used in Fig.~\ref{brentjens_fig:rmtf_0}}
\label{brentjens_fig:sigma_rm}
\end{figure}

\begin{figure*}
\centering
\includegraphics[width=\textwidth]{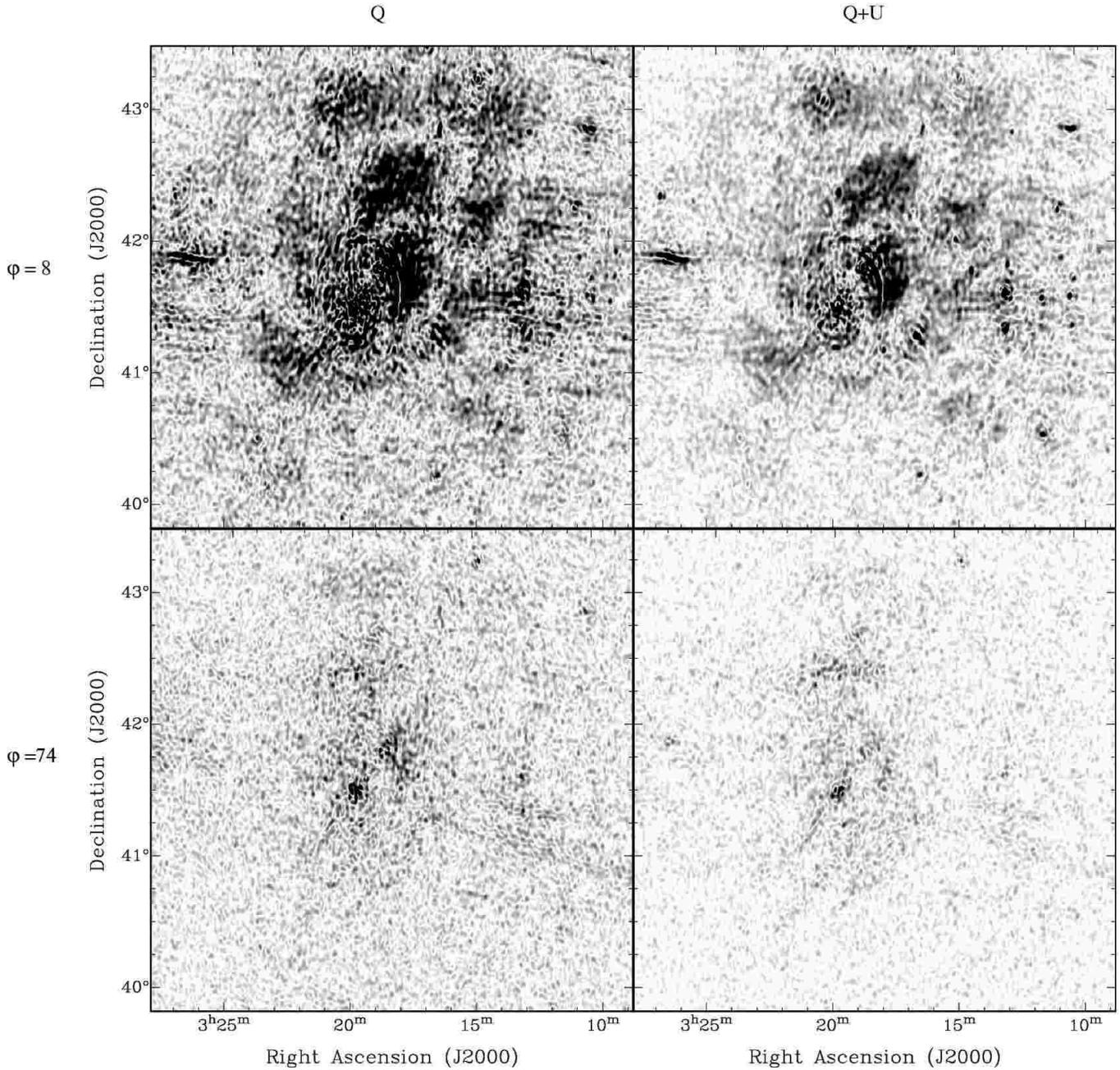}
\caption{Comparison of $Q$-only RM-synthesis (left) and $Q+U$
RM-synthesis (right) for Faraday depths $\phi = 8\ \mbox{rad}\
\mbox{m}^{-2}$ (top) and $\phi = 74\ \mbox{rad}\ \mbox{m}^{-2}$
(bottom). The panels show total linear polarization in grey scale. The
scale runs from 0.15 mJy~beam$^{-1}$~rmtf$^{-1}$ to 1.5
mJy~beam$^{-1}$~rmtf$^{-1}$. }
\label{brentjens_fig:rmsynthesis_perseus}
\end{figure*}

The standard error in the RM when obtained by fitting a straight line
to a plot of $\chi$ versus $\lambda^2$ is given by
\begin{eqnarray}
\label{brentjens_eqn:sigma_rm}
\sigma_\mathrm{RM} & = &
\frac{\sigma_\chi}{\sigma_{\lambda^2}\sqrt{N-2}}\\
\sigma_{\lambda^{2}} &=& \sqrt{\left(N-1\right)^{-1}\sum_{i=1}^{N}\lambda_i^4 - \lambda_0^4},
\end{eqnarray}
where $\lambda_0$ is given by equation
(\ref{brentjens_eqn:average_lambda2}) and $N$ is the number of
channels. $\sigma_\chi$ is either
\begin{equation}
\sigma_\mathrm{\chi}  = \frac{1}{2}\frac{\sigma}{\|P\|}
\end{equation}
if $\sigma_\mathrm{Q} \approx \sigma_\mathrm{U} = \sigma$ or
\begin{equation}
\sigma_\mathrm{\chi}  = \frac{\sqrt{U^2\sigma_\mathrm{Q}^2 +
Q^2\sigma_\mathrm{U}^2}}{2\|P\|^2}
\end{equation}
if $\sigma_\mathrm{Q}$ and $\sigma_\mathrm{U}$ differ by more than a
factor of two or so. $\sigma_{\lambda^{2}}$ is the standard deviation
of the distribution of $\lambda^2_i$ points. This is given by the
frequency setup of the instrument. It is a measure of the effective
width of the $\lambda^2$ distribution. Equation
(\ref{brentjens_eqn:sigma_rm}) is the result of straightforward error
propagation. For reference, a full derivation is presented in appendix
\ref{brentjens_sec:sigma_rm}.

In RM-synthesis, one determines the RM of a single source along the
line of sight by fitting, for example, a parabola to the main peak of
$\|F(\phi)\|$. The detailed procedure is to first find the brightest
point in a critically sampled Faraday dispersion function (2--3 points
per $\delta\phi$), covering a wide range in $\phi$. This is followed
by oversampling the region around the peak by a large factor. A
parabolic fit to the 10--20 points directly surrounding the peak then
yields the RM of the source.

We have simulated this procedure in order to get a quantitative idea
of the typical error in RM that one obtains, given a certain noise
level in the Stokes $Q$ and $U$ images, and a certain set of sample
points $\lambda^2_i$. The results are shown in
Fig.~\ref{brentjens_fig:sigma_rm}. The total signal-to-noise ratio is
equal to $\sqrt{N-2}\,\|P\|/\sigma$. The solid line is equation
(\ref{brentjens_eqn:sigma_rm}). The points are standard deviations in
RM computed from 1000 iterations per S/N ratio point. We have assumed
the noise in Stokes $Q$ and $U$ to be equal and Gaussian. We see
excellent agreement with the error expected for traditional
$\lambda^2$ fitting (the straight line). At a S/N ratio less than 4,
the points deviate strongly from the line. This is due to the fact
that the non-Gaussianity of the noise in $P$ is only noticeable close
to the origin of the complex plane. It is stressed that a total S/N of
4 when having 126 channels implies a S/N per channel of slightly less
than 0.4. It is impossible to determine a polarization
angle with such a low S/N in the case of standard $\lambda^2$ fitting.

\section{RM-synthesis with only $Q$ or $U$}
\label{brentjens_sec:qu_only}

It is also possible to perform a RM-synthesis with Stokes $Q$ or $U$
only. There exist many radio observations that have produced only
Stokes $I$ and $Q$, for example spectral line work with arrays
equipped with linearly polarized feeds, or data from backends with
limited correlator capacity. However, by using only one of the two
Stokes parameters, one loses information about the sign of the Faraday
depth.

The derivation is started with equation
(\ref{brentjens_eqn:inversion}). The identities
\begin{eqnarray}
Q & = & (P + P^\ast)/2 \label{brentjens_eqn:Q}\\
\mathrm{i}U & = & (P - P^\ast)/2 \label{brentjens_eqn:iU}
\end{eqnarray}
and
\begin{eqnarray}
G(x) & = & \int_{-\infty}^{\infty} g(t)
\mathrm{e}^{a\mathrm{i}xt}\mathrm{d}t \Leftrightarrow
G^\ast(-x) =  \int_{-\infty}^{\infty} g^\ast(t)
\mathrm{e}^{a\mathrm{i}xt}\mathrm{d}t \label{brentjens_eqn:conjugate_ft}.
\end{eqnarray}
are needed. After substituting  equations (\ref{brentjens_eqn:Q}) and
(\ref{brentjens_eqn:iU}) into equation
(\ref{brentjens_eqn:inversion}) and using equation
(\ref{brentjens_eqn:conjugate_ft}), one obtains:
\begin{eqnarray}
\frac{1}{2}\left(\tilde{F}(\phi) + \tilde{F}^\ast(-\phi) \right)& = & K\int_{-\infty}^{+\infty}\tilde{Q}
(\lambda^2)\mathrm{e}^{-2\mathrm{i}\phi (\lambda^2-\lambda_0^2)}\ 
\mathrm{d}\lambda^2\\
\frac{1}{2}\left(\tilde{F}(\phi) - \tilde{F}^\ast(-\phi) \right)& = & K\int_{-\infty}^{+\infty}\mathrm{i}\tilde{U}
(\lambda^2)\mathrm{e}^{-2\mathrm{i}\phi (\lambda^2-\lambda_0^2)}\ 
\mathrm{d}\lambda^2.
\end{eqnarray}

The sensitivity to emission at low Faraday depth (less than one full
rotation over the entire band) is limited by the orientation of the
polarization vector. If the emission is mostly in $U$ at $\lambda =
\lambda_0$, and hardly in $Q$, one will not retrieve the full total
polarization when using Stokes $Q$ images only. On the other hand, if
all emission is in $Q$, one apparently retrieves twice the actual
total polarized intensity. Both cases are usually not very
relevant, because RM-synthesis is mostly applied when emission at high
Faraday depth is expected, where the polarization vector makes several
turns over the observed band.

Fig.~\ref{brentjens_fig:rmsynthesis_perseus} compares
results of a complete RM-synthesis of data of the Perseus cluster,
taken with the WSRT \citep{DeBruynBrentjens2005}, to results of a
$Q$-only RM-synthesis of the same dataset. It compares both the
Galactic foreground emission at low Faraday depth, and the emission at
higher Faraday depth that we attribute to the Perseus cluster. It is
clearly seen that the noise in the $Q$-only images is increased with
respect to the complete RM-synthesis. The bar-like feature at
$\alpha\approx 3^\mathrm{h}18^\mathrm{m}$, $\delta\approx +42\degr
30\arcmin$ is already visible in the $Q$-only images. This
demonstrates that one actually can detect faint emission at high
Faraday depths using only Stokes $Q$ or $U$ images. Unless the
situation is simple, meaning only one discrete source along the line
of sight, these images are unfortunately not useful in a quantitative
sense. However, it is an efficient way to discover weak, Faraday
rotated, polarized emission in existing datasets, which can
then be followed up with full polarization observations.

\section{General experiment layout}
\label{brentjens_sec:experiment_layout}

\begin{figure}
\resizebox{\hsize}{!}{\includegraphics{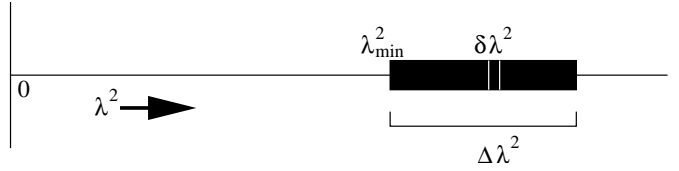}}
\caption{The three instrumental parameters that determine the output
of a Faraday rotation experiment.} 
\label{brentjens_fig:threeparameters}
\end{figure}

Three main parameters are involved when planning a rotation-measure
experiment, namely the channel width $\delta\lambda^2$, the width of
the $\lambda^2$ distribution $\Delta\lambda^2$, and the shortest
wavelength squared $\lambda^2_\mathrm{min}$. They are summarized in
Fig.~\ref{brentjens_fig:threeparameters}. These parameters determine
respectively the maximum observable Faraday depth, the resolution in
$\phi$ space, and the largest scale in $\phi$ space to which one is
sensitive.  Estimates for the FWHM of the main peak of the RMTF, the
scale in $\phi$ space to which sensitivity has dropped to 50\% and
the maximum Faraday depth to which one has more than 50\% sensitivity
are approximately
\begin{eqnarray}
\delta\phi &\approx& \frac{2\sqrt{3}}{\Delta\lambda^2}
\label{brentjens_eqn:rmtf_width}\\
\mbox{max-scale} &\approx&\frac{\pi}{\lambda^2_\mathrm{min}}
\label{brentjens_eqn:max_scale}\\
\|\phi_\mathrm{max}\| &\approx& \frac{\sqrt{3}}{\delta\lambda^2}.
\label{brentjens_eqn:phi_max}
\end{eqnarray}
In these equations we assumed a top hat weight
function which is 1 between $\lambda^2_\mathrm{min}$ and
$\lambda^2_\mathrm{max}$ and zero elsewhere.

It is interesting to compare equations
(\ref{brentjens_eqn:rmtf_width}) and (\ref{brentjens_eqn:max_scale}).
This is where the analogy between RM-synthesis and regular synthesis
imaging breaks down. In synthesis imaging, the width of the
synthesized beam is inversely proportional to the maximum absolute
$uv$ vector. That is, the distance between the origin and the $uv$
point most distant from it. The maximum scale that one can measure
depends on the shortest baseline. Therefore one is always
maximally sensitive to structures smaller than the width of the
synthesized beam.

This is quite different in RM-synthesis. In RM-synthesis it
is possible that a source is unresolved in the sense that its extent
in $\phi$ is less than the width of the RMTF, yet "resolved" out
because one has not sampled the typical $\phi$-scale of the source due
to lack of small $\lambda^2$ points. Equation
(\ref{brentjens_eqn:rmtf_width}) shows that the width of the RMTF
depends on the width of the $\lambda^2$ distribution, not on the
largest $\lambda^2$ measured. Nevertheless the largest scale in $\phi$
that one is sensitive to is set by the smallest $\lambda^2$ as is
shown in equation (\ref{brentjens_eqn:max_scale}).  In order to truly
resolve Faraday thick clouds in $\phi$ space in the sense that one
could see internal structure, the main peak of the RMTF should be
narrower than the maximum scale to which one is sensitive. Because
$2\sqrt{3} \approx \pi$, the requirement for resolving Faraday thick
structures is
\begin{equation}
\lambda^2_\mathrm{min} < \Delta\lambda^2.
\label{brentjens_ref:resolve_condition}
\end{equation}
See appendix \ref{brentjens_sec:example_simulations} for
simulations illustrating this point.

For deconvolution the RMTF should be known as accurately as possible
for all sources within the field of view and along the line of
sight. The main problems are:
\begin{itemize}
\item frequency dependence of the primary beam attenuation;
\item frequency dependence of the instrumental polarization;
\item the intrinsic emission spectra of the sources;
\item frequency dependence of the synthesized beam size.
\end{itemize}

The last point is easily compensated for by using the same $uv$-taper
for all channel maps. The first two points can be alleviated to a
large extent by observing in mosaic mode. The primary beam attenuation
can cause very steep artificial spectral indices. These can be as
steep as $\alpha \approx -1.5$ near half power and $\alpha \approx -5$
near the $-10$~dB point at the WSRT. Fortunately, the primary beam
attenuation is usually known accurately. Therefore one can predict
$W(\lambda^2)$ for any point in the field of view, enabling one to
accurately compute the local RMTF at any pixel using for example
equation (\ref{brentjens_eqn:rmtf_sum}).

After primary beam correction, one should align the channel maps
spectrally. Our preferred method is to determine the average
total intensity of a large sample of sources, and scale the images until
the average of the ensemble in a particular channel map matches the
value at $\lambda = \lambda_0$. Using this approach, the
spectra of as yet undetected emission should be approximately flat to
within a spectral index range of $\pm 1$. Of course one could flatten
source spectra on an individual basis. This is only useful if one is
interested in bright sources that are easily detected in individual
channels.

A convenient property of RM-synthesis is that more-or-less frequency
independent instrumental problems end up at $\phi = 0$,
convolved with the RMTF. This means that instrumental problems are
highly reduced at higher absolute Faraday depths. In other words: at
high Faraday depth, we ``wind-up'' the instrumental polarization
problems, while ``unwinding'' the Faraday rotated cosmic polarization
signals.

\section{Conclusions}
\label{brentjens_sec:conclusions}

We have extended the work of \citet{Burn1966} to the cases of limited
sampling of $\lambda^2$ space and some spectral dependencies. We have
introduced the RMTF, which is an excellent predictor of $n\pi$
ambiguity problems in the frequency setup. RM-synthesis can be
implemented very efficiently on modern computers. For example, a RM-
synthesis of 126 input maps of 1024$^2$ pixels, yielding $3\times100$
output maps of $1024^2$ pixels ($P$, $Q$, and $U$) takes less than 5
minutes on a laptop equipped with a 2 GHz Intel Pentium processor and
512 MB of RAM.

Because the analysis is easily applied to wide fields, one can conduct
very fast RM surveys of weak sources. Difficult situations, for
example multiple sources along the line of sight, are easily
detected. Under certain conditions, it is even possible to recover the
emission as a function of Faraday depth within a single cloud of
ionized gas.

Instrumental problems that are weakly frequency independent, or
have a very characteristic frequency dependence, are easily separated
from cosmic signals that are only subject to Faraday rotation.

Rotation measure synthesis has already been successful in discovering
widespread, weak, polarized emission associated with the Perseus cluster
\citep{DeBruynBrentjens2005}. In simple, high signal to noise
situations it is as good as traditional linear fits to $\chi$ versus
$\lambda^2$ plots. However, when the situation is more complex, or
very weak polarized emission at high rotation measures is expected, it
is the only viable option.

\begin{acknowledgements}
We acknowledge Robert Braun, Torsten En\ss lin and Peter Katgert for
useful and vivid discussions on the subject. The Westerbork Synthesis
Radio Telescope is operated by ASTRON (Netherlands Foundation for
Research in Astronomy) with support from the Netherlands Foundation
for Scientific Research (NWO).
\end{acknowledgements}

\appendix
\section{Standard errors in RM estimations}
\label{brentjens_sec:sigma_rm}

The expected standard errors in RM/Faraday depth and $\chi_0$ are
useful quantities when planning a rotation measure experiment. In this
appendix we present a formal derivation. 

From  equations (\ref{brentjens_eqn:p_exp}) and
(\ref{brentjens_eqn:p_qiu}) we have
\begin{eqnarray}
\|P\| &=& \sqrt{Q^2+U^2}\\
\chi  &=& \frac{1}{2}\tan^{-1}\frac{U}{Q}.
\end{eqnarray}

We discriminate two cases. The first is $\sigma_\mathrm{Q} \approx
\sigma_\mathrm{U} = \sigma$, the second is $\sigma_\mathrm{Q} \gg
\sigma_\mathrm{U}$ or $\sigma_\mathrm{Q} \ll \sigma_\mathrm{U}$, where
$\gg$ and $\ll$ indicate a difference of more than a factor of
two. $\sigma_\mathrm{Q}$ and $\sigma_\mathrm{U}$ are the RMS image
noise in individual $Q$ and $U$ channel maps. 

The derivation is done in two steps. First we derive the standard
error in the polarization angle and total polarization, $\sigma_\chi$
and $\sigma_\mathrm{P}$ of measurements in individual
channels. Then we apply standard results for the least squares fit of
a straight line to obtain $\sigma_\phi$ and $\sigma_{\chi_0}$, the
standard errors in rotation measure / Faraday depth and the
polarization angle at $\lambda = 0$.

Error propagation \citep{Squires} gives us
\begin{eqnarray}
\sigma_\mathrm{P}^2 & = & \left(\frac{\partial \|P\|}{\partial
Q}\right)^2\sigma_\mathrm{Q}^2 + \left(\frac{\partial \|P\|}{\partial
U}\right)^2\sigma_\mathrm{U}^2 \label{brentjens_eqn:sigma_p2}\\
\sigma_\chi^2 & = & \left(\frac{\partial\chi}{\partial
Q}\right)^2\sigma_\mathrm{Q}^2 + \left(\frac{\partial \chi}{\partial
U}\right)^2\sigma_\mathrm{U}^2\label{brentjens_eqn:sigma_chi2}.
\end{eqnarray}

The partial derivatives for $\sigma_\mathrm{P}$ are
\begin{eqnarray}
\left(\frac{\partial \|P\|}{\partial
Q}\right)^2 & = & \frac{Q^2}{Q^2+U^2}\label{brentjens_eqn:dpdq}\\
\left(\frac{\partial \|P\|}{\partial
U}\right)^2 & = & \frac{U^2}{Q^2+U^2}.\label{brentjens_eqn:dpdu}
\end{eqnarray}

Inserting this in equation (\ref{brentjens_eqn:sigma_p2}) gives
\begin{equation}
\sigma_\mathrm{P}^2  =\frac{Q^2}{\|P\|^2}\sigma_\mathrm{Q}^2
+\frac{U^2}{\|P\|^2}\sigma_\mathrm{U}^2.\label{brentjens_eqn:sigma_p2_b}
\end{equation}
In the most general case, this cannot be simplified any
further. Nevertheless, when $\sigma_\mathrm{Q} \approx
\sigma_\mathrm{U} = \sigma$, equation (\ref{brentjens_eqn:sigma_p2_b})
can be simplified to
\begin{equation}
\sigma_\mathrm{P}^2  = \sigma^2.
\label{brentjens_eqn:sigma_p2_c}
\end{equation}

The partial derivatives needed for $\sigma_\chi^2$ are
\begin{eqnarray}
\frac{\partial \frac{1}{2}\tan^{-1} \frac{U}{Q}}{\partial Q} & = &
 \frac{1}{4}\frac{U^2}{\left(Q^2 + U^2\right)^2}\\
\frac{\partial \frac{1}{2}\tan^{-1} \frac{U}{Q}}{\partial U} & = &
 \frac{1}{4}\frac{Q^2}{\left(Q^2 + U^2\right)^2}.
\end{eqnarray}

Inserting these in equation (\ref{brentjens_eqn:sigma_chi2}) gives
\begin{equation}
\sigma_\mathrm{\chi}^2  = \frac{U^2\sigma_\mathrm{Q}^2 +
Q^2\sigma_\mathrm{U}^2}{4\|P\|^4}
\end{equation}

This result can only be simplified further if $\sigma_\mathrm{Q} \approx
\sigma_\mathrm{U} = \sigma$:
\begin{equation}
\sigma_\mathrm{\chi}^2  = \frac{1}{4}\frac{\sigma^2}{\|P\|^2}.
\label{brentjens_eqn:sigma_chi2_a}
\end{equation}

When fitting a straight line $y = ax+b$ to data with equal estimated
standard errors per data point, the standard error in the slope of the
line is
\citep{Squires}
\begin{equation}
\sigma_\mathrm{a}^2 \approx \frac{1}{N-2} \frac{\Sigma_i \left(y_i -
ax_i-b\right)^2}{\Sigma_i x_i^2 - N^{-1}\left(\Sigma_i x_i\right)^2}.
\end{equation}
Now one substitutes $\phi = a$, $\chi_0 = b$, $\chi = y$, and
$\lambda^2 = x$. A rotation measure of 0 may be assumed without loss
of generality. The equation then becomes
\begin{equation}
\sigma^2_\phi = \frac{1}{N-2}\frac{\Sigma_i
\chi_i^2}{\Sigma_i\left(\lambda^2_i\right)^2 - N^{-1}\left(\Sigma_i
\lambda^2_i\right)}
\label{brentjens_eqn:sigma_phi2_a}
\end{equation}
The variance of the polarization angle distribution is given by
\begin{equation}
\sigma_\chi^2 = \frac{1}{N-1}\Sigma_i \chi_i^2 - \langle\chi\rangle,
\label{brentjens_eqn:sigma_chi2_b}
\end{equation}
where $\langle\rangle$ denotes averaging.  Combining equations
(\ref{brentjens_eqn:sigma_phi2_a}) and
(\ref{brentjens_eqn:sigma_chi2_b}) and using $\langle\chi\rangle = 0$
one obtains
\begin{equation}
\sigma^2_\phi =
\frac{N-1}{N-2}\frac{\sigma_\chi^2}{\Sigma\left(\lambda^2_i\right)^2 -
N^{-1}(\Sigma_i \lambda_i^2)^2}
\label{brentjens_eqn:sigma_phi2_b}
\end{equation}

The variance of the $\lambda^2$ distribution is
\begin{equation}
\sigma_{\lambda^2}^2 = \frac{1}{N-1}\left(\Sigma_i \lambda_i^4 -
N^{-1}\left(\Sigma_i \lambda_i^2\right)^2\right).
\label{brentjens_eqn:sigma_l2}
\end{equation}
This is a measure of the effective width of the $\lambda^2$
coverage.

Equation (\ref{brentjens_eqn:sigma_phi2_b}) may now be simplified by
substituting equations (\ref{brentjens_eqn:sigma_l2}) and
(\ref{brentjens_eqn:sigma_chi2_a}). The final result is:
\begin{equation}
\sigma_\phi^2 = \frac{\sigma^2}{4 \left(N-2\right)\|P\|^2\sigma^2_{\lambda^2}}.
\end{equation}

The standard error in in $\chi_0$, $\sigma_{\chi_0}$ is derived in a
similar fashion. One starts with
\begin{equation}
\sigma^2_\mathrm{b} \approx \left(\frac{\Sigma_i x_i^2 -
N^{-1}\left(\Sigma x_i\right)^2}{N} + \langle
x\rangle^2\right)\sigma^2_\mathrm{a}.
\end{equation}
After substituting equations (\ref{brentjens_eqn:sigma_l2}) and
(\ref{brentjens_eqn:average_lambda2}) and some
rearranging, the final result is:
\begin{equation}
\sigma^2_{\chi_0} =
\frac{\sigma^2}{4\left(N-2\right)\|P\|^2}\left(\frac{N-1}{N} +
\frac{\lambda_0^4}{\sigma_{\lambda^2}^2}\right)
\end{equation}


\section{Example simulations}
\label{brentjens_sec:example_simulations}

\begin{figure*}[p]
\centering
\includegraphics[width=\textwidth]{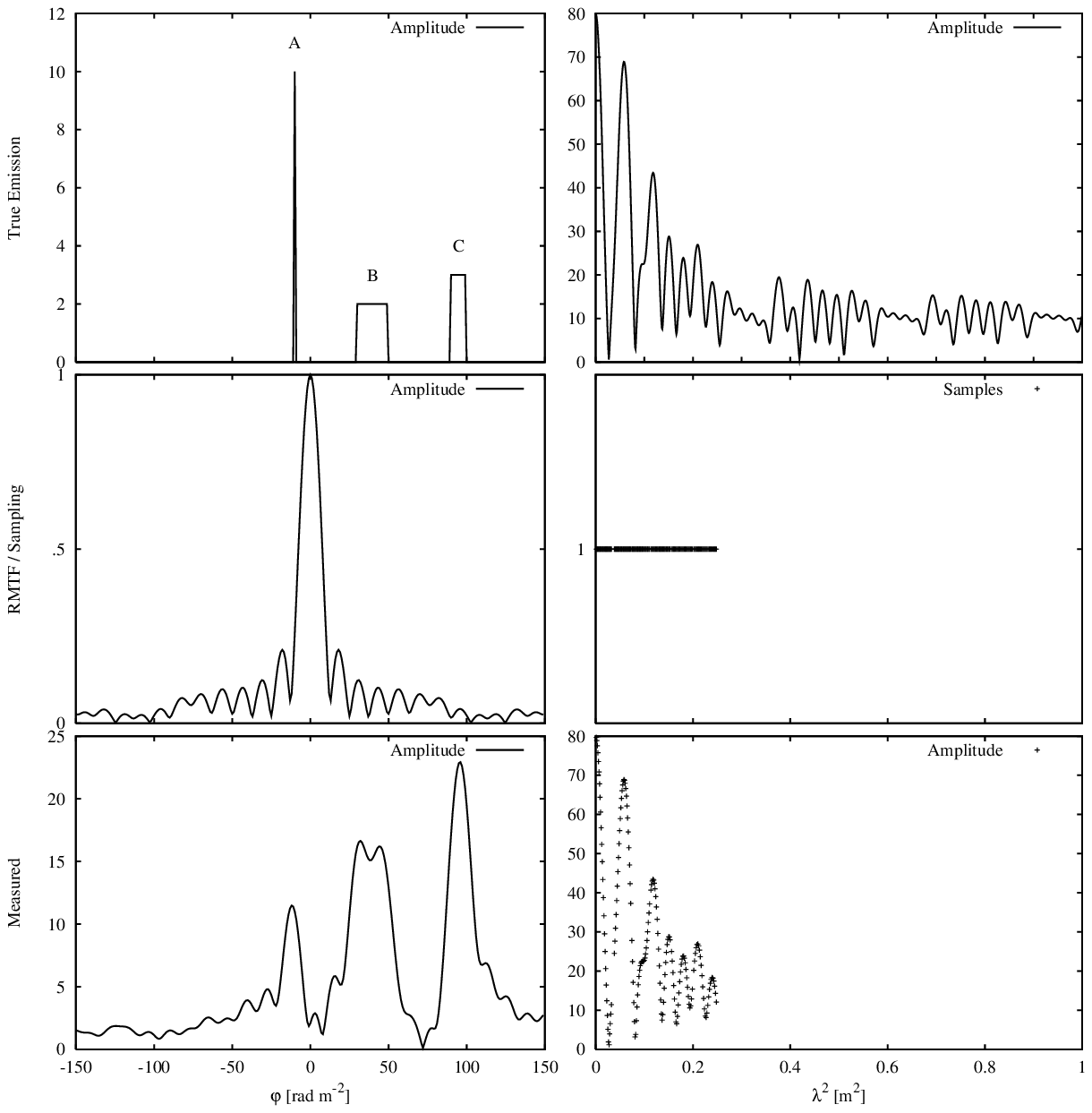}
\caption{Wavelength range: $3.6$--$50\ \mbox{cm}$.}
\label{brentjens_fig:extended_short}
\end{figure*}

\begin{figure*}[p]
\centering
\includegraphics[width=\textwidth]{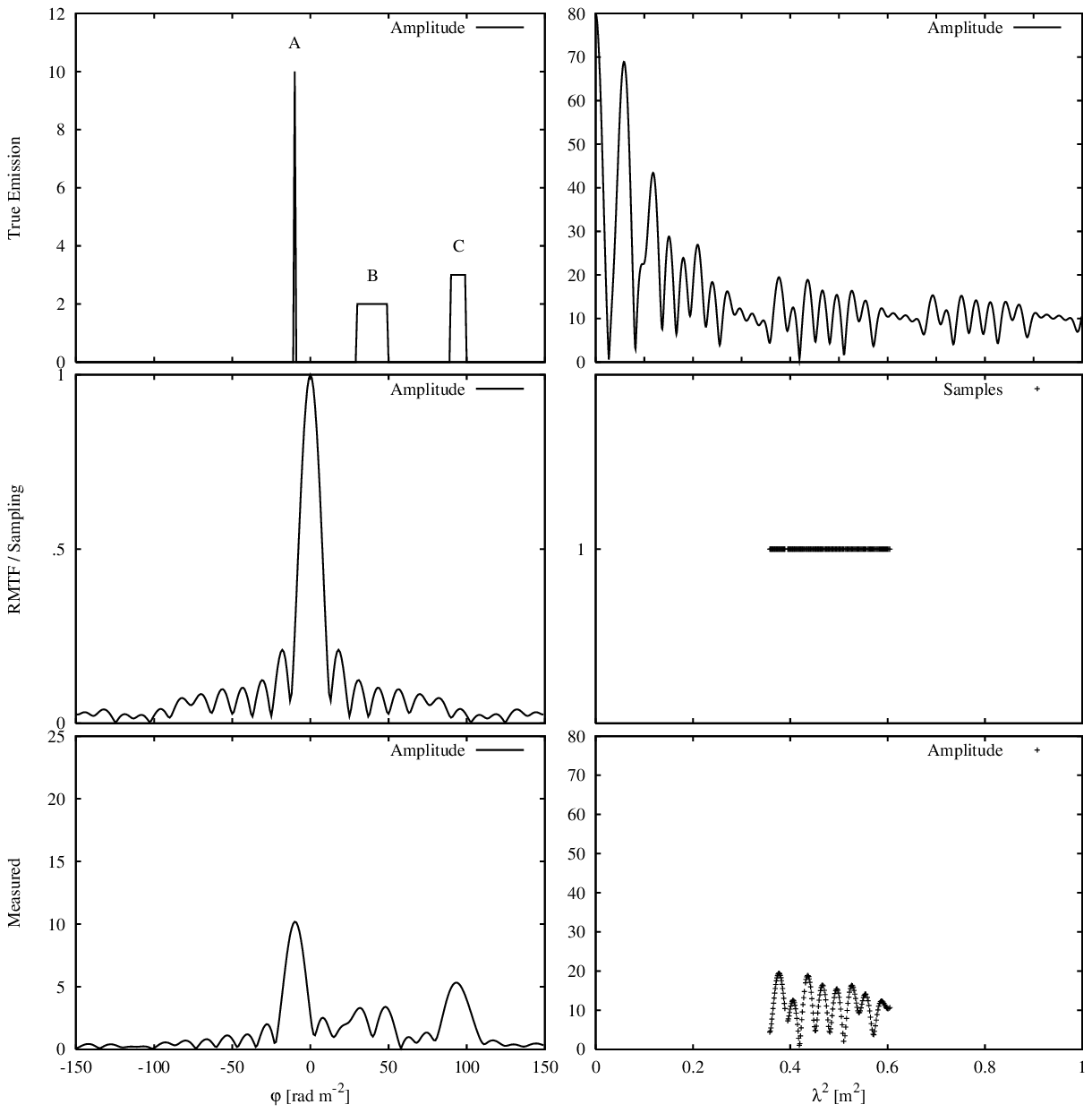}
\caption{Wavelength range: $60$--$78\ \mbox{cm}$.}
\label{brentjens_fig:extended_medium}
\end{figure*}

\begin{figure*}[p]
\centering
\includegraphics[width=\textwidth]{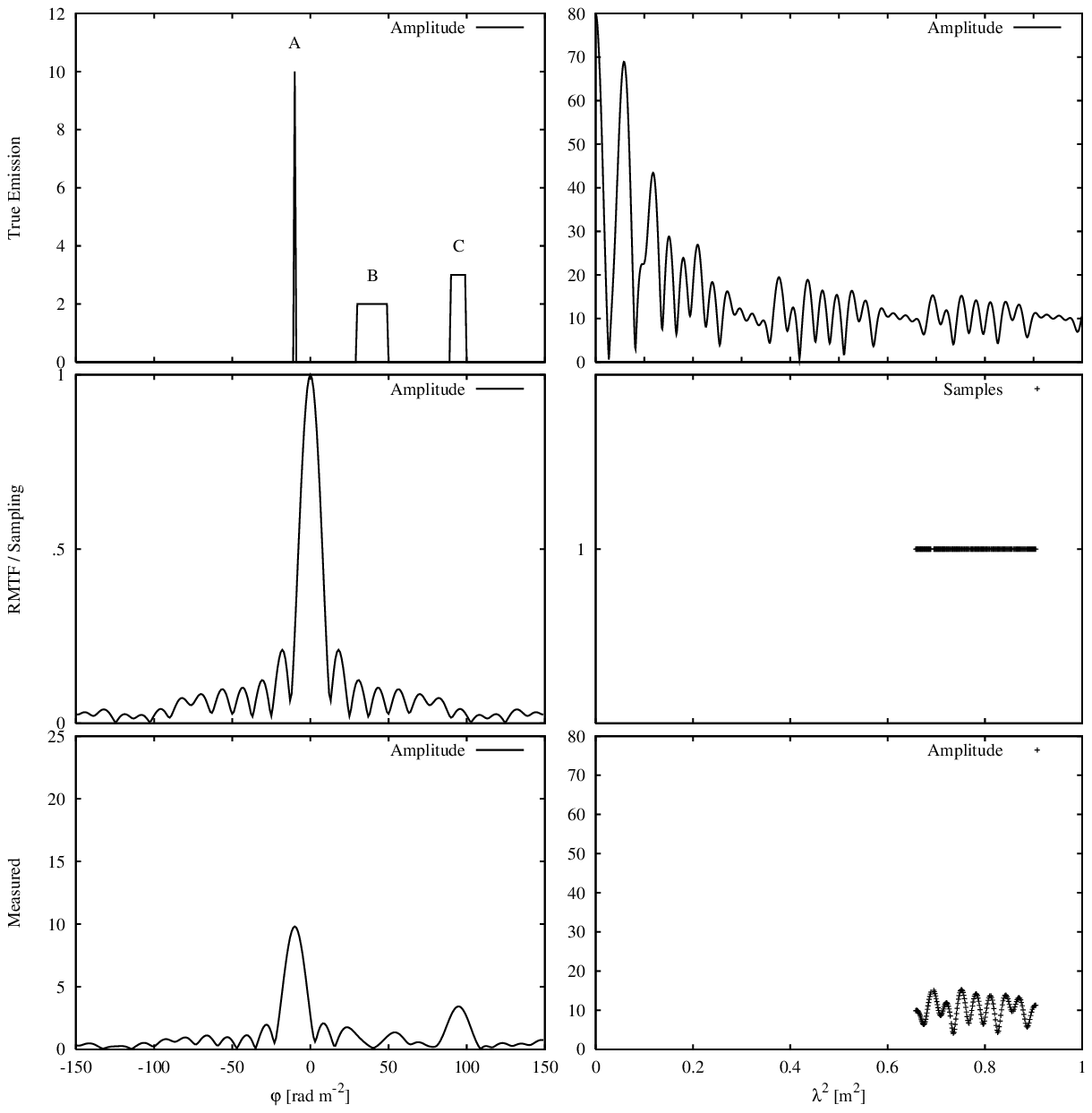}
\caption{Wavelength range: $81$--$95\ \mbox{cm}$.}
\label{brentjens_fig:extended_long}
\end{figure*}

In this appendix we show, as an illustration, three model runs of an
RM-synthesis of an artificial Faraday dispersion function, measured
with a realistic frequency sampling. We hope that these figures aid in
understanding the most important aspects of RM-synthesis specifically
and rotation measure work in general.

Sources that are extended in the plane of the sky have their surface
brightness measured in Jy per steradian. For point sources the flux in Jy
is sufficient to characterize it. The respective brightness units for
sources that are both extended in the plane of the sky and in Faraday
depth are Jy steradian$^{-1}$ (rad m$^{-2}$)$^{-1}$ or Jy m$^{2}$
rad$^{-3}$. Sources that are extended in the plane of the sky and
point-like in $\phi$ space have their brightness in $\phi$ space
measured in Jy steradian$^{-1}$. The brightness of the measured
Faraday dispersion function has units of Jy (beam on the sky)$^{-1}$
(rmtf)$^{-1}$. Sources that are point-like in the plane of the sky
have the steradian$^{-1}$ or (beam on the sky)$^{-1}$ removed.

In order to keep the units simple, we made all simulated sources
point-like in the sky plane. Hence the units used in the figures in
this appendix are:
\begin{itemize}
\item top right and bottom right: Jy
\item middle left and middle right: dimensionless
\item top left: Jy m$^2$ rad$^{-1}$
\item bottom left: Jy rmtf$^{-1}$
\end{itemize}

The lefthand column of Fig.~\ref{brentjens_fig:extended_short},
Fig.~\ref{brentjens_fig:extended_medium}, and
Fig.~\ref{brentjens_fig:extended_long} is the situation in $\phi$
space. The righthand column is the corresponding situation in
$\lambda^2$ space. Pictures in one row are converted into each other
by Fourier relations (\ref{brentjens_eqn:p_observed_fourier}) (left to
right) and (\ref{brentjens_eqn:inversion}) (right to left). The top
row is the input situation. It is the same in all three figures. The
lefthand panel represents the polarized flux per unit Faraday depth in
Jy per (rad m$^{-2}$). The righthand panel is the polarized flux in
Jy. The example sources all have flat spectra. The second row shows
the RMTF in the lefthand panel and the sampling function in the
righthand panel. The third row shows the result of applying the second
row to the first row. That is, multiplication of top right and middle
right give bottom right and top left convolved with middle left after
$\lambda^2$ bandpass filtering gives bottom left. Except for the
weight of the samples in $\lambda^2$ space, all quantities are
complex. For clarity, we have only shown the amplitude here.

The RMTF in all three figures is the same because the pattern and
width of the $\lambda^2$ coverage is exactly the same for all of
them. The only difference is the absolute position of the
pattern. Fig.~\ref{brentjens_fig:extended_short} has 
$\lambda^2_\mathrm{min} = 3.6^2\ \mbox{cm}^2$,
Fig.~\ref{brentjens_fig:extended_medium} has $\lambda^2_\mathrm{min} =
60^2\ \mbox{cm}^2$, and Fig.~\ref{brentjens_fig:extended_long} has
$\lambda^2_\mathrm{min} = 81^2\ \mbox{cm}^2$.

The three sources in this simulation have different properties to
illustrate different cases.
\begin{description}
\item[A:] $\phi = -10\ \mbox{rad}\ \mbox{m}^{-2}$, delta function of
$\phi$, total flux density is 10 Jy. In synthesis imaging, this would
be the equivalent of a point source;
\item[B:] $30 \le \phi \le 50\ \mbox{rad}\ \mbox{m}^{-2}$, multiple
RMTFs wide in $\phi$, $F(\phi) = 2\ \mbox{Jy}\ \mbox{m}^2\
\mbox{rad}^{-1}$, total flux density is 40 Jy. In synthesis imaging,
this would be the equivalent of an extended source;
\item[C:] $90 \le \phi \le 100\ \mbox{rad}\ \mbox{m}^{-2}$, roughly
one RMTF wide in $\phi$, $F(\phi) = 3\ \mbox{Jy}\ \mbox{m}^2\
\mbox{rad}^{-1}$, total flux density is 30 Jy. In synthesis imaging,
this would be the equivalent of a barely resolved source.
\end{description}

Because source A is a delta function with respect to $\phi$,
the amplitude of its Fourier transform is the same at all
$\lambda^2$. Therefore, the source appears in all three figures with
equal amplitude. The peak is slightly higher than 10 Jy rmtf$^{-1}$ in
Fig.~\ref{brentjens_fig:extended_short} and
Fig.~\ref{brentjens_fig:extended_medium} due to side lobes of the
responses of sources B and C. In
Fig.~\ref{brentjens_fig:extended_long} the retrieved flux of sources B
and C has collapsed so much that the response to source A is
practically the same as the RMTF, especially to the left of source A.

Source B represents the other extreme. Being several RMTFs wide, one
requires $\lambda^2_\mathrm{min} \ll \Delta\lambda^2$ in order to
recover the full flux of the source. Only
Fig.~\ref{brentjens_fig:extended_short} meets this requirement. In
Fig.~\ref{brentjens_fig:extended_medium}, only two bumps at the edges
of the source remain. Because in
Fig.~\ref{brentjens_fig:extended_medium} we only sample smaller scales
in $\phi$ due to the larger $\lambda^2_\mathrm{min}$, the only parts
of source B that remain are the parts where these smaller scales are
important: the edges. Source B has practically disappeared in
Fig.~\ref{brentjens_fig:extended_long}.

Source C is of an intermediate type. Because its typical $\phi$-scale is
narrower than source B, there is a larger fraction of the total flux
recovered in Fig.~\ref{brentjens_fig:extended_medium} and
Fig.~\ref{brentjens_fig:extended_long}.

In analogy to radio interferometric observations, one could state that
the $\lambda^2$ sampling in
Fig.~\ref{brentjens_fig:extended_short} corresponds to a connected
element array, where one samples all scales up to
$\lambda^2_\mathrm{max}$ approximately equally well.
Fig.~\ref{brentjens_fig:extended_long} corresponds to a VLBI
observation, where one misses the short spacings and therefore is
insensitive to extended emission. A fundamental difference with radio
interferometry is that the resolution in $\phi$ space is determined by
the width of the $\lambda^2$ distribution, $\Delta\lambda^2$, and not
by the largest $\lambda^2$ sampled. Hence one could encounter
situations where a source is \emph{not} resolved in the sense that the
thickness of the source in $\phi$ is much less than the width of the
RMTF, while at the same time it \emph{is} resolved out in the sense
that one has not sampled sufficiently short $\lambda^2$ points to
detect the source.

\bibliographystyle{aa}
\bibliography{2990}

\end{document}